# Oscillating Detonation of Liquid Ammonia

Wenhao Wang[a,b,c], Zongmin Hu[a,c,*], and Peng Zhang[b,*]

[a] *State Key Laboratory of High-Temperature Gas Dynamics (LHD), Institute of Mechanics, Chinese Academy of Sciences, Beijing 100190, China*

[b] *Department of Mechanical Engineering, City University of Hong Kong, Kowloon Tong, Kowloon 999077, Hong Kong*

[c] *School of Engineering Sciences, University of Chinese Academy of Sciences, Beijing 100049, China*

**Abstract:**

Liquid ammonia is a promising carbon-free energy carrier, but its high volatility and low reactivity lead to detonation dynamics that differs significantly from those of liquid hydrocarbons. Using Eulerian–Lagrangian simulations, we revealed an oscillating detonation phenomenon driven by the flash boiling of ammonia. Specifically, intense endothermic evaporation and exothermic combustion periodically weaken and then restore the coupling between the shock front and the reaction zone. A delay differential equation (DDE) model is developed to describe this oscillatory behaviour. In this model, the shock-response constraint is derived from the positive characteristic compatibility relation behind the shock front, and it is closed using delay-augmented relaxation models for evaporation and reaction progress variables. Normal-mode stability analysis of the DDE system shows that an oscillatory solution emerges when the evaporation and reaction timescales are comparable. Simulation data across different evaporation models, droplet diameters, and ambient temperatures collapse onto the theoretical frequency band predicted by the model.

## 1. Introduction

Detonation represents a canonical mode of supersonic combustion, characterised by a leading shock front immediately followed by a reaction zone [1-3]. Its potential for near-constant-volume heat release and high thermodynamic efficiency has motivated extensive research in pulsed [4], rotating [5], and oblique detonation wave engines [6]. In the context of global carbon neutrality, ammonia ($NH_3$) has emerged as a prominent zero-carbon fuel candidate due to its high hydrogen density and established infrastructure [7]. For practical propulsion utility, ammonia is predominantly stored and injected in its liquid state [8]. This urges us to consider ammonia detonation as a two-phase reacting flow problem: the leading shock is intimately coupled with following droplet breakup, vaporisation, mixing, and reaction [9-11]. Particularly, the stability of ammonia detonation waves—where weak chemical reactivity competes with rapid phase-change and interfacial processes—remains important yet insufficiently understood.

Theoretical inquiry into the stability of gaseous detonation spans over a century. Although the classical Zeldovich-von Neumann-Doring (ZND) model offers a mathematically steady, planar wave structure [12], experiments and numerical simulations consistently demonstrate that such a structure is intrinsically unstable. A critical feature of the ZND structure is the existence of a sonic locus that isolates the reaction zone from downstream perturbations, a mechanism that renders the detonation self-sustained [12, 13]. Seminal normal-mode linear stability analyses, pioneered by Erpenbeck [14, 15], revealed that the stability of the ZND structure is determined by the spectrum of eigenvalues of the linearised disturbance equations. Bourlioux, Majda, and Roytburd [16] further suggested that the extreme temperature sensitivity of the induction zone, particularly in the long-wavelength limit, renders the large-activation-energy asymptotic solution physically unrealisable. This theoretical landscape was later mapped with much greater precision by Lee and Stewart [17], who developed a robust shooting technique to compute continuous neutral stability curves and eigenfunctions. By imposing an acoustic-radiation condition at the downstream boundary, they extended this shooting-based linear stability analysis to the Chapman–Jouguet (CJ) limit. This shooting methodology was later generalised by Sharpe [18] to classify the unstable spectrum over a range of reaction orders.

Abouseif and Toong [19] combined numerical simulations with a linearised stability theory based on the high-activation-energy approximation, showing that interactions between entropy–temperature fluctuations and the finite reaction zone generates an oscillatory energy-source field. The phenomenological understanding was placed on a rigorous mathematical footing by Clavin and co-workers. Clavin and He [20] derived a nonlinear integral equation based on a quasi-isobaric approximation for overdriven detonations, explicitly identifying the instability mechanism as a resonant coupling between the leading shock and acoustic modes within the reaction zone. This was further generalised by Clavin, He, and Williams [21] to include multidimensional perturbations and later refined by Daou and Clavin [22] to determine the precise instability thresholds for ordinary detonations, confirming that the coupling of acoustic modes with the reaction rate is the primary destabilising factor.

In the nonlinear regime, the hydrodynamic instability manifests in rich dynamical behaviours ranging from periodic pulsations to chaos. Kasimov, Faria, and Rosales [23] identified the heat of reaction and activation energy as the primary controlling parameters of the bifurcation from regular to chaotic structures. Short and Quirk [24] investigated chain-branching kinetics, identifying the cross-over temperature as a bifurcation parameter governing the route to period-doubling and quenching. Bridging linear theories and simulations, Kasimov and Stewart [25] derived asymptotic evolution equations describing the nonlinear dynamics. Furthermore, Ng et al. [26] and Abderrahmane, Paquet and Ng [27] utilised bifurcation theory and power spectral density analysis to show that the birth of higher-order instability modes follows the classic Feigenbaum scenario, marking a route to chaos via period-doubling. Accurately capturing these dynamics poses significant numerical challenges. As noted by Romick, Aslam and Powers [28], although shock-fitting methods provide spectral accuracy for linear modes, high-resolution shock-capturing schemes are essential to retrieve the transient features of weakly and mildly unstable detonations. Beyond instantaneous dynamics, the mean structure of unstable detonations has become a focal point of modern research. Radulescu et al. [29] and Sow, Chinnayya, and Hadjadj [30] investigated the hydrodynamic structure of cellular detonations and proposed that hydrodynamic fluctuations behave analogously to turbulent eddies. In their formulation, mechanical fluctuations act as an additional exothermic process, effectively modifying the sonic condition and extending the reaction zone length; the wave thus propagates with a reaction-zone length that exceeds the steady ZND value.

Realistic detonation wave propagation involves boundary and real-gas effects. Zhang and Lee [31, 32] demonstrated that frictional momentum losses can paradoxically destabilise a wave—mimicking an increase in activation energy. However, Dionne, Ng and Lee [33] clarified that excessive drag ultimately leads to failure. At the diffusive scale, Han, Wang, and Law [34] and Mazaheri, Mahmoudi and Radulescu [35] highlighted the critical role of molecular transport in dampening high-frequency modes and regularising the reaction zone. Moreover, Wang and Mével [36] addressed high-pressure regimes, utilising the Noble-Abel equation of state to quantify the effects of finite molecular volume on stability through both linear analysis and simulation.

The presence of a dispersed phase in the form of liquid fuel droplets fundamentally alters the stability characteristics of the gaseous detonation established in the above theories. The propagation of a detonation wave through a droplet-laden mixture involves a hierarchy of strongly coupled processes: shock-induced droplet breakup, interphase momentum transfer, fuel evaporation, and chemical reactions [9]. Lu and Law [37] elucidated the competition between evaporative cooling and chemical heat release and showed that the global reaction rate can transition from an Arrhenius-controlled to a vaporisation-controlled regime. Tricard and Zhao [38] showed that low-volatility fuels are more prone to entering the vaporisation-controlled regime, thereby increasing their sensitivity to quenching caused by heat and momentum deficits. This two-way coupling fundamentally modifies the detonation wave structure. Martínez-Ruiz [39] revealed that the interphase exchange of mass and energy within the post-shock relaxation zone alters the Chapman–Jouguet (CJ) conditions relative to the gaseous limit. Hernández-Sánchez, Huete, and Ruiz [40] further demonstrated that the endothermic nature of vaporisation can lead to “pathological” detonations, where the sonic locus shifts to an intermediate position within the reaction zone. Macroscopically, these exchanges manifest as velocity deficits. Kailasanath [11] highlighted that a critical droplet size threshold (typically about 10 μm) is required to sustain near-ideal CJ velocities. Cheatham and Kailasanath [41] corroborated that the deviation from the gaseous limit is primarily driven by mixture inhomogeneity resulting from finite vaporisation rates, rather than the specifics of interphase transport sub-models.

The presence of droplets also influences the stability of detonation front. Short and Stewart [9] formulated a stability framework for condensed-phase ZND detonations and obtained the neutral stability boundaries through a normal-mode analysis. The discrete phase interacts with the cellular structure as the transverse waves entrain particles along triple-point trajectories. Watanabe et al. [10, 42] identified it as the precursor to a downstream, pseudo-turbulent dispersive regime. Meng et al. [43] characterised this aerodynamic scattering as a source of vapour heterogeneity, where the local antagonism between shock compression and evaporative heat absorption dictates extinction or re-initiation dynamics. Furthermore, Schwer [44] identified a direct coupling between the macroscopic cell size and the microscopic vaporisation timescale, noting that persistent pockets of unreacted fuel in two-dimensional simulations exacerbate velocity deficits compared to one-dimensional approximations. Yet, the dispersed phase does not solely destabilise the front; Huang et al. [45] showed that the interplay between finite disturbances and evaporative enrichment can damp longitudinal pulsations, thereby suppressing instabilities.

Although the fundamental mechanisms of spray detonations are well-charted for hydrocarbon fuels, research on ammonia detonation remains in its infancy, primarily due to its distinct physicochemical properties. Past efforts have largely focused on chemical sensitisation to overcome high initiation thresholds. Zhu et al. [7, 46] identified the hydrogen blending ratio as the governing parameter for the deflagration-to-detonation transition (DDT) and velocity deficits, effectively accelerating global reaction kinetics. Zhu et al. [47] extended their work to droplet-laden flows, identifying that droplet size and number density are critical parameters controlling extinction limits. Sun et al. [48, 49] examined cracked ammonia-air mixtures, revealing that low cracking ratios necessitate significantly higher initiation energies and proposing an indirect initiation strategy that utilises the transition zone to stabilise propagation. Veiga-López et al. [50] observed a trade-off in ammonia-hydrogen-air blends: higher hydrogen content reduces entropy production but exacerbates $NO_x$ emissions.

The existing literature reveals a conspicuous scarcity of research addressing the hydrodynamic instability of liquid ammonia detonations, particularly concerning the impact of phase change. Liquid ammonia is distinguished by its high volatility, contrasting sharply with the low-volatility fuels (e.g., Decane and Dodecane) that underpin standard quasi-steady evaporation models. Under the extreme thermodynamic forcing behind a detonation shock, liquid ammonia is prone to flash boiling—a non-equilibrium phase transition. How this rapid vapour generation interacts with detonation remains little understood. The present study bridges this gap through a combined computational and theoretical investigation, focusing specifically on the regime where the conventional assumption of timescale separability collapses—namely, when the flash boiling rate becomes comparable to the chemical reaction rate. Our computations revealed critical dynamic behaviours ranging from initiation failure to a distinct oscillating instability mode driven by fast evaporation, characterized by the recurrently weakened and strengthened coupling between the shock front and the reaction zone. To formulize these understandings, we developed a normal-mode stability analysis of a delay differential equation (DDE) system to demonstrate that the breakdown of timescale separability fundamentally reconfigures neutral stability boundaries. This theory collapses the simulated frequencies across evaporation model, droplet diameter, and ambient temperature, and shows that phase change can act as an active stability-selection mechanism in liquid-ammonia detonation dynamics.

The rest of this paper is organised as follows. The governing equations, physical sub-models, and computational specifications are detailed in Section 2, accompanied by grid and parcel convergence verification. Section 3 presents the results and discussion. The structural bifurcation induced by flash boiling and the resulting oscillating instability mechanism are characterised in Section 3.1, which also examines the parametric sensitivity of the oscillation dynamics. Section 3.2 presents a DDE model and validates the derived dispersion relations against the numerical data. Finally, concluding remarks are provided in Section 4.

## 2. Formulation and methodology

### *2.1 Governing equations*

This study uses an Eulerian–Lagrangian framework to computationally investigate detonation in dispersed two-phase flows. The evolution of the continuous phase is governed by the multi-component compressible Navier–Stokes equations,

$$\frac{\partial \rho}{\partial t} + \nabla \cdot (\rho \boldsymbol{u}) = S_m, \tag{2.1}$$

$$\frac{\partial (\rho \boldsymbol{u})}{\partial t} + \nabla \cdot [\boldsymbol{u}(\rho \boldsymbol{u})] + \nabla p - \nabla \cdot \boldsymbol{\tau} = \boldsymbol{S}_F, \tag{2.2}$$

$$\frac{\partial (\rho E)}{\partial t} + \nabla \cdot [\boldsymbol{u}(\rho E + p)] + \nabla \cdot \boldsymbol{q} - \nabla \cdot (\boldsymbol{\tau} \cdot \boldsymbol{u}) = S_e, \tag{2.3}$$

$$\frac{\partial (\rho Y_i)}{\partial t} + \nabla \cdot [\boldsymbol{u}(\rho Y_i)] + \nabla \cdot [-D\nabla(\rho Y_i)] = \dot{\omega}_i + S_{s,i}, (i = 1,2, \dots, ns - 1). \tag{2.4}$$

Here, the variables $\rho$, $\boldsymbol{u}$, and $p$ represent the gas density, velocity, and pressure, respectively. System closure requires the ideal gas law, $p = \rho RT$, where $T$ is the gas temperature and $R$ is the specific gas constant calculated as $R = R_u \sum_{i=1}^{ns}(Y_i/MW_i)$. Here, $R_u = 8.314\ \mathrm{J/(mol \cdot K)}$ is the universal gas constant, and $MW_i$ is the molecular weight of the $i$-th species. The number of species is $ns$, and $Y_i$ is the mass fractions of $i$-th species. The net chemical production rate of species $i$ is $\dot{\omega}_i$. Chemical kinetics is modelled by a skeletal mechanism comprising 34 species and 204 elementary reactions proposed by Song et al. [51]. This mechanism has been validated extensively against experimental benchmarks and accurately reproduces fundamental detonation characteristics in ammonia-fuelled systems [7, 46]. The deviatoric stress tensor is $\boldsymbol{\tau} = \mu[\nabla \boldsymbol{u} + (\nabla \boldsymbol{u})^T - 2(\nabla \cdot \boldsymbol{u})\mathbf{I}/3]$, where $\mu$ is the dynamic viscosity calculated by using Sutherland's law. The total energy is $E = e + |\boldsymbol{u}|^2/2$, where $e$ is the specific internal energy. The diffusive heat flux follows Fourier's law, $\boldsymbol{q} = -k\nabla T$, with thermal conductivity $k$. $D$ is the density-weighted diffusion coefficient obtained under the unity-Lewis-number assumption as $D = k/(\rho C_p)$, where $C_p$ is the heat capacity at constant pressure given by $C_p = \sum_{i=1}^{ns} Y_i C_{p,i}$. Finally, the source terms $S_m$, $\boldsymbol{S}_F$, $S_e$, and $S_{s,i}$ account for the interphase exchange of mass, momentum, energy, and species with the dispersed droplet phase, the formulations of which are detailed below.

The dispersed phase is evolved within a Lagrangian framework, where the mass, velocity, and temperature of individual droplets are governed using Lagrangian particle tracking (LPT) method by the following system of ordinary differential equations

$$\frac{dm_d}{dt} = \dot{m}_d, \tag{2.5}$$

$$\frac{d\boldsymbol{u}_d}{dt} = \frac{\boldsymbol{F}_d}{m_d}, \tag{2.6}$$

$$c_l \frac{dT_d}{dt} = \frac{\dot{Q}_c + \dot{Q}_l}{m_d}, \quad (2.7)$$

where $m_d$, $\boldsymbol{u}_d$, and $T_d$ are the droplet mass, velocity, and temperature. The droplet mass is calculated as $m_d = \pi \rho_l D_d^3/6$ under the assumption that the droplet is spherical, where $\rho_l$ and $D_d$ are the liquid density and droplet diameter. The specific heat capacity of the liquid at constant pressure is denoted by $c_l$. The driving terms on the right-hand side—representing the rates of mass exchange ($\dot{m}_d$), aerodynamic forcing ($\boldsymbol{F}_d$), and convective and latent heat transfer ($\dot{Q}_c$, $\dot{Q}_l$)—are determined by the constitutive sub-models detailed below.

Two evaporation models were employed in the present study. The first is the conventional one proposed by Abramzon and Sirignano [52]. Ignoring the singular dynamics of boiling, it establishes a lower bound on the vaporisation rate for volatile liquid. Under the quasi-steady assumption, the droplet mass depletion rate is expressed as:

$$\dot{m}_d = -\pi D_d Sh \overline{D}_f \rho_s \, ln(1 + B_M), \quad (2.8)$$

where $\overline{D}_f$ is the binary diffusion coefficient averaged over the gas film surrounding the droplet. The gas density at the droplet surface, $\rho_s$, is determined via the equation of state $\rho_s = p_s MW_m/(R_u T_s)$, where $p_s$ is evaluated using a polynomial function of $T_s$ [53], and the reference film temperature $T_s$ is approximated using the "one-third rule", $T_s = (T + 2T_d)/3$. The convective enhancement of interfacial mass transfer is modelled by the Ranz–Marshall correlation, $Sh = 2.0 + 0.6 Re_d^{1/2} Sc^{1/3}$, where *Sh* is the Sherwood number. The droplet Reynolds number, $Re_d = \rho_g D_d |\boldsymbol{u}_d - \boldsymbol{u}|/\mu$, is based on the gas-phase properties and the slip velocity, and $Sc = \nu/\overline{D}_f$ is the Schmidt number, with $\nu$ being the kinematic viscosity of gas. The Spalding mass transfer number is $B_M = (Y_s - Y_\infty)/(1 - Y_s)$, where $Y_\infty$ is the fuel vapour mass fraction in the far field (surrounding gas) and $Y_s$ is the vapour mass fraction at the droplet interface. The latter is computed by $Y_s = MW_f X_s / \sum_i (X_i MW_i)$, where $MW_f$ is the molecular weight of the fuel vapour and $X_s$ represents the surface mole fraction of the fuel vapour. Assuming local thermodynamic equilibrium for species at the liquid–gas interface, the interfacial vapour mole fraction is evaluated from Raoult's law as $X_s = X_l p_{\text{sat}}(T_d)/p$, where $X_l$ is the liquid mole fraction of the condensing species and $p_{\text{sat}}(T_d)$ is the saturation pressure, evaluated as a polynomial function of the droplet temperature $T_d$ [53].

Due to the low boiling point of liquid ammonia, the droplet evaporation under the condition of detonation may be dominated by flash boiling, which are inherently transient and non-equilibrium. To capture this physics, we also employed the model of Zuo et al. [54] for $T_d \geq T_{\text{sat}}(p)$. This model accounts for both nucleation and flash boiling, thereby yielding significantly higher mass transfer rates. The total evaporation rate is decomposed into superheated ($\dot{m}_{sh}$) and subcooled ($\dot{m}_{sc}$) components:

$$\dot{m}_d = \dot{m}_{sc} + \dot{m}_{sh}. \quad (2.9)$$

The superheat component is driven by the excess thermal energy within the liquid and modelled by Adachi's correlation [55]:

$$\dot{m}_{sh} = \frac{\alpha_s (T_d - T_b) A_d}{L_v(T_b)}, \quad (2.10)$$

where $A_d = \pi D_d^2$ is the droplet surface area, $T_b$ is the liquid boiling temperature, and $L_v(T_b)$ is the latent heat at $T_b$. The overall heat transfer coefficient $\alpha_s$ (in kW m$^{-2}$ K$^{-1}$) is evaluated via the piecewise function:

$$\alpha_s = \begin{cases} 0.76(T_d - T_b)^{0.26}, & and \quad 0 < T_d - T_b < 5 \\ 0.027(T_d - T_b)^{2.33} and, & 5 < T_d - T_b < 25 \\ 13.8(T_d - T_b)^{0.39}, & T_d - T_b \geq 25 \end{cases} \tag{2.11}$$

The subcooled evaporation rate $\dot{m}_{sc}$ satisfied

$$\dot{m}_{sc} = \frac{2\pi k D_d Nu}{C_p(1 + \dot{m}_{sh}/\dot{m}_{sc})} \ln\left[1 + \left(1 + \frac{\dot{m}_{sh}}{\dot{m}_{sc}}\right)\frac{h_\infty - h_b}{L(T_b)}\right], \tag{2.12}$$

where $h_\infty$ and $h_b$ denote the specific enthalpy of the gas mixture in the far field and at the droplet surface, respectively. This flash-boiling formulation has been extensively validated and demonstrated to accurately reproduce the evaporation of ammonia under high-pressure and high-temperature conditions [56, 57].

The interphase aerodynamic coupling is governed by the drag force $\boldsymbol{F}_d$ of spherical particles:

$$\boldsymbol{F}_d = -\left(\frac{18\mu}{\rho_l D_d^2}\right)\left(\frac{C_d Re_d}{24}\right) m_d(\boldsymbol{u}_d - \boldsymbol{u}), \tag{2.13}$$

where $C_d$ is the drag coefficient. For the high Reynolds number regimes typical of detonation flows ($\mathrm{Re}_d > 1000$), we adopted $C_d = 0.424$ [58]. Zhu et al. [59] demonstrated that this model introduces a maximum deviation of only 6.4% compared with more complex correlations (e.g., Brown–Lawler), confirming its sufficiency for capturing momentum exchange under supersonic conditions.

Interphase energy transfer comprises both convective heating and latent cooling. The convective heat transfer rate $\dot{Q}_c$ is given by

$$\dot{Q}_c = h_c A_d (T - T_d) \tag{2.14}$$

where $h_c = Nuk/D_d$ is the convective heat transfer coefficient. The Nusselt number $Nu$ is evaluated with the Ranz-Marshall correlation [60] as $Nu = 2.0 + 0.6Re_d^{1/2}Pr^{1/3}$, where $Pr = c_p\mu/k$ is the gas-phase Prandtl number. The latent-heat term is written $\dot{Q}_l = -\dot{m}_d h_l(T_d)$, where $h_l(T_d)$ is the latent heat of vaporisation evaluated at the droplet temperature $T_d$. The Ranz–Marshall correlation is a standard choice in two-phase detonation solvers and has been widely employed and validated in related two-phase detonation studies [59, 61-63].

The Pilch–Erdman model was used for droplet breakup [64], and it has been extensively applied in liquid-fuelled detonation simulations [43, 65-68]. The model divides the breakup process into five experimentally identified regimes, each associated with a different characteristic breakup time. The corresponding Weber number range spans from $O(10)$ to $O(10^3)$, covering the conditions relevant to the present study.

To resolve the interphase interactions discussed above, we implemented the Particle-Source-In-Cell (PSI-CELL) method [69]. The gas-phase conservation equations are modified by source terms derived from the Lagrangian parcels within each computational cell of volume $V_c$. Summing over the $N_p$ droplet parcels present in the cell, the source terms for mass, momentum, energy, and species transport are:

$$S_m = -\frac{1}{V_c}\sum_1^{N_p} \dot{m}_d\,, \tag{2.15}$$

$$\boldsymbol{S}_F = -\frac{1}{V_c}\sum_1^{N_p} \boldsymbol{F}_d\,, \tag{2.16}$$

$$S_e = -\frac{1}{V_c}\sum_{1}^{N_p}\left(\dot{Q}_c + \dot{Q}_l\right), \tag{2.17}$$

$$S_{s,i} = -\frac{1}{V_c}\sum_{1}^{N_p}\dot{m}_{d,i}\,. \tag{2.18}$$

In (2.18), the subscript "$i$" denotes only condensed phases and $\dot{m}_{d,i}$ is the corresponding evaporation rate.

*2.2 Computational methodology and validation*

The governing equations described in Section 2.1 were solved using an open-source compressible reacting-flow solver developed by the present authors within the OpenFOAM V7 framework [70]. This solver extends the *rhocentralFoam* algorithm by incorporating detailed chemical kinetics, multi-component transport, and Lagrangian particle tracking modules. The gas-phase equations are discretised using the finite-volume method. Time integration is performed with a first-order Euler scheme, spatial gradients are evaluated using the "Gauss-Limited Linear" scheme, and convective fluxes are computed using the Kurganov–Noelle–Petrova (KNP) central-upwind scheme to ensure robust shock capturing. The Courant–Friedrichs–Lewy (CFL) number is restricted to 0.05 throughout the simulations.

The solver has been validated against one- and two-dimensional benchmarks for liquid-fuelled detonations [71, 72]. To further establish its fidelity for ammonia detonation, the computed wave velocities for gaseous ammonia–oxygen mixtures are compared with the corresponding theoretical Chapman–Jouguet (CJ) values over a range of global equivalence ratios, as summarised in table 1. The agreement is consistently good, with discrepancies below 2% for all cases considered. Since the CJ velocity provides a fundamental integral measure of detonation propagation, this comparison confirms that the solver captures the principal thermochemical and compressible-flow features relevant to the present study.

| **Equivalence ratio** | **Theory [m/s]** | **Present study [m/s]** | **Error** |
|---|---|---|---|
| 0.6 | 2210.89 | 2243.21 | 1.4 % |
| 0.8 | 2344.75 | 2379.97 | 1.5 % |
| 1.0 | 2437.63 | 2481.20 | 1.8 % |
| 1.2 | 2506.61 | 2540.65 | 1.4 % |
| 1.5 | 2543.97 | 2584.24 | 1.6 % |

Table 1. Validation against theoretical Chapman–Jouguet (CJ) predictions for the gaseous ammonia–oxygen detonation speeds across a range of equivalence ratios ($\phi$).

*2.3 Computational configuration*

The present study considers the transmission of a steady gaseous detonation from an upstream vapour zone into a downstream droplet-laden region. Figure 1 illustrates the configuration: an already-formed steady gaseous detonation responds to a downstream droplet-laden field. An upstream gaseous section (0.5 m long) contains a 10 mm ignition zone followed by a premixed ammonia-vapour region. This gaseous driver section transitions downstream into a 1.0 m test section containing oxygen seeded with monodisperse liquid-ammonia droplets. The droplets (diameter, $D_d$ = 20 or 40 μm) are initialised at 233 K and distributed uniformly to yield a global equivalence ratio of unity ($\phi$ = 1). To explore sensitivity to ambient conditions, the initial gas temperature in the test section is set to either 300 K or 220 K. The left boundary is modelled as a rigid reflecting wall, while the right boundary employs a zero-gradient supersonic outflow condition.

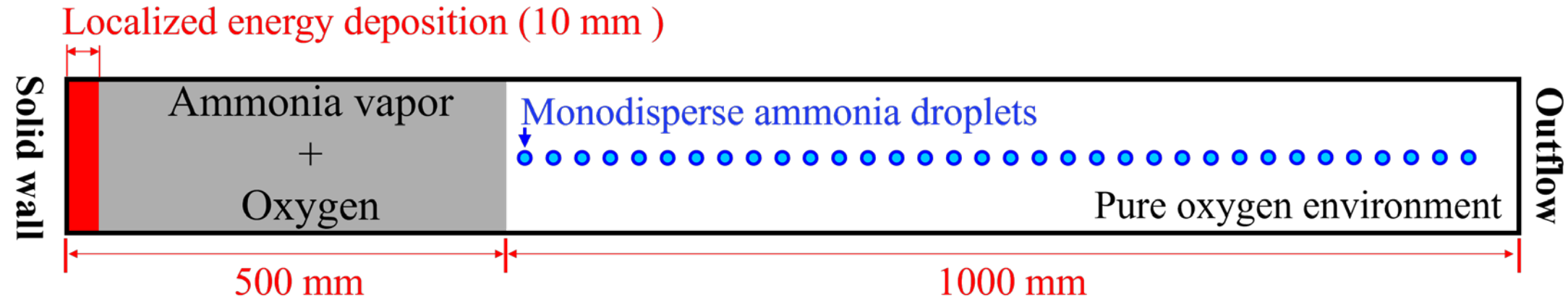

Fig. 1 One-dimensional computational configuration. A 0.5 m gaseous driver section generates a steady CJ detonation that transmits into a 1.0 m droplet-laden test section ($O_2$ at 300 K or 220 K and 71 kPa; monodisperse $NH_3$ droplets at 233 K, $D_d$ = 20 μm or 40 μm, $\phi$ = 1). The left boundary is a solid wall and the right boundary is a zero-gradient supersonic outflow; the detonation is initiated by a 10 mm localized energy deposition (2500 K, $7.1\times10^7$ Pa) placed adjacent to the left wall inside the gaseous driver section.

## *2.4 Grid and parcel convergence*

To ensure numerical fidelity, convergence tests were performed for a representative stable -propagation baseline case introduced in Section 2.3, with respect to both the spatial resolution and the number of Lagrangian parcels. Figure 2 shows the instantaneous spatial distributions of pressure, temperature and $H_2O$ mass fraction. The left column (Figure 2(a)) examines grid convergence for mesh sizes of 0.20, 0.10 and 0.05 mm, while the right column (Figure 2(b)) assesses the sensitivity to the total number of Lagrangian parcels, namely $1.5 \times 10^4$, $3.0 \times 10^4$ and $6.0 \times 10^4$. In both cases, the solutions exhibit clear convergence as the numerical resolution is increased. For the grid study, only minor differences remain near the detonation front, mainly between the 0.20 mm mesh and the two finer meshes, whereas the results obtained with 0.10 and 0.05 mm are nearly indistinguishable. For the parcel-number study, the insets provide a magnified view of the wavefront and show excellent agreement between the results obtained with $3.0 \times 10^4$ and $6.0 \times 10^4$ parcels. This indicates that the adopted grid and parcel resolutions are sufficient for the present simulations.

Table 2 reports the key detonation parameters for both convergence studies. The detonation velocity ($U_D$) exhibits exceptional stability, varying by less than 0.1% between the 0.10 mm and 0.05 mm resolutions. Likewise, the peak von Neumann pressures ($P_{VN}$) differ by no more than 0.2 atm, and the Chapman-Jouguet temperature ($T_{CJ}$) shows only slight sensitivity. When combined with the profiles shown in Figure 2(a), this close agreement among mesh sizes suggests that the 0.05 mm grid provides a balance between computational expense and solution accuracy, thereby justifying its selection for all the subsequent simulations.

Turning to the influence of particle resolution, similarly, all three key detonation metrics exhibit a high degree of consistency across the cases. This stability indicates that a parcel count of $3.0 \times 10^4$ sufficiently captures the multiphase interactions without imposing excessive computational costs. When considered alongside the profiles in Figure 2(b), the results demonstrate a clear convergence with respect to parcel number, confirming its negligible impact on simulation outcomes. Consequently, a parcel number of $3.0 \times 10^4$ is adopted for all subsequent calculations.

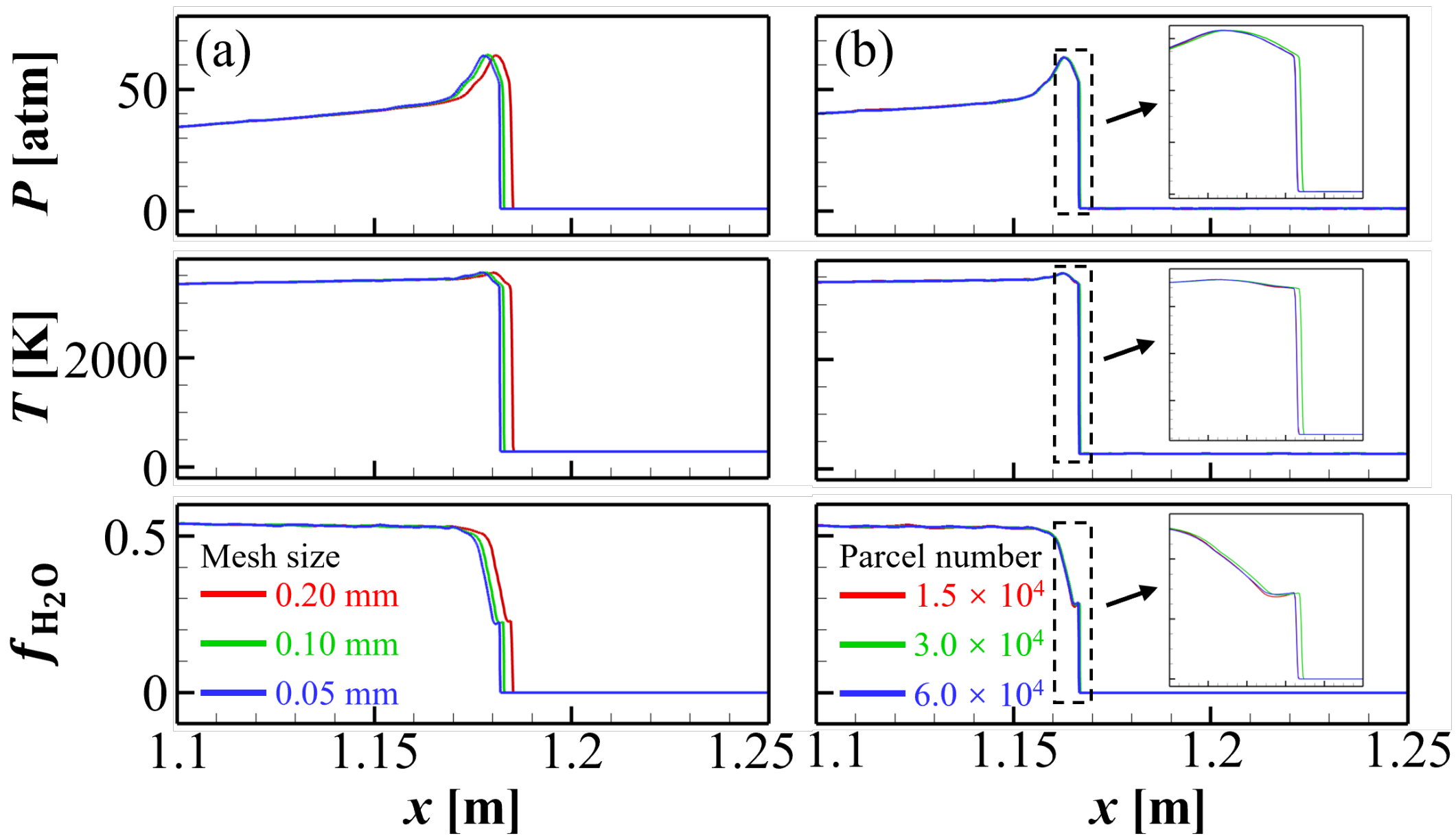

Fig.2 Convergence study at $t = 0.5$ms for a baseline case in which the detonation wave propagates steadily through the droplet-laden mixture without flash boiling $\left(T_{O_2} = 300 \text{ K}, D_d = 20 \text{ µm}\right)$. Profiles of pressure, temperature, and $H_2O$ mass fraction are compared. (a) Mesh-dependence test using three grid spacings, $\Delta x = 0.20$, $0.10$, and $0.05$ mm. (b) Parcel-dependence test using three parcel numbers, $N_p = 1.5 \times 10^4$, $3.0 \times 10^4$, and $6.0 \times 10^4$. The complementary parameter is held fixed in each test: $N_p = 3.0 \times 10^4$ in (a) and $\Delta x = 0.10$mm in (b). Insets magnify the shock front. The two finer levels are nearly indistinguishable in both tests; therefore, $\Delta x = 0.05$ mm and $N_p = 3.0 \times 10^4$ are adopted in the production simulations and used for the configuration shown in Fig. 1.

| Case | Variation | $U_D$ [m/s] | $P_{VN}$ [atm] | $T_{CJ}$ [K] |
|---|---|---|---|---|
| $N_p = 3\times10^4$ | $\Delta x = 0.20$ mm | 2450.55 | 63.87 | 3457.17 |
| | $\Delta x = 0.10$ mm | 2449.57 | 63.98 | 3456.40 |
| | $\Delta x = 0.05$ mm | 2449.47 | 64.11 | 3446.68 |
| $\Delta x = 0.10$ mm | $N_p = 1.5\times10^4$ | 2452.75 | 64.12 | 3475.34 |
| | $N_p = 3.0\times10^4$ | 2449.47 | 64.11 | 3446.68 |
| | $N_p = 6.0\times10^4$ | 2451.08 | 63.42 | 3472.69 |

Table 2. Comparison of detonation velocity ($U_D$), von Neumann pressure ($P_{VN}$), and CJ temperature ($T_{CJ}$) for various mesh and parcel numbers.

## 3. Results and Discussion

### *3.1 Oscillating detonation of liquid ammonia*

#### *3.1.1 Flash-boiling-induced pressure oscillation*

Figure 3 shows the pressure evolution of the reference liquid-ammonia two-phase detonation case, with initial gas temperature 300 K and droplet diameter 20 μm. The initially overdriven detonation wave relaxes into a steady leading front in the upstream vapour zone, as indicated by the nearly constant $P_{\text{max}}$. Once the detonation wave enters the droplet-laden region, its subsequent evolution depends strongly on droplet evaporation. Without flash boiling, the detonation remains stable and $P_{\text{max}}$ rises to a higher value. With flash boiling, the detonation wave undergoes a transition from stable to oscillatory propagation, and $P_{\text{max}}$ develops pronounced periodic fluctuations. The detailed thermochemical structure responsible for this phenomenon is examined below using the instantaneous wave profiles and species distributions.

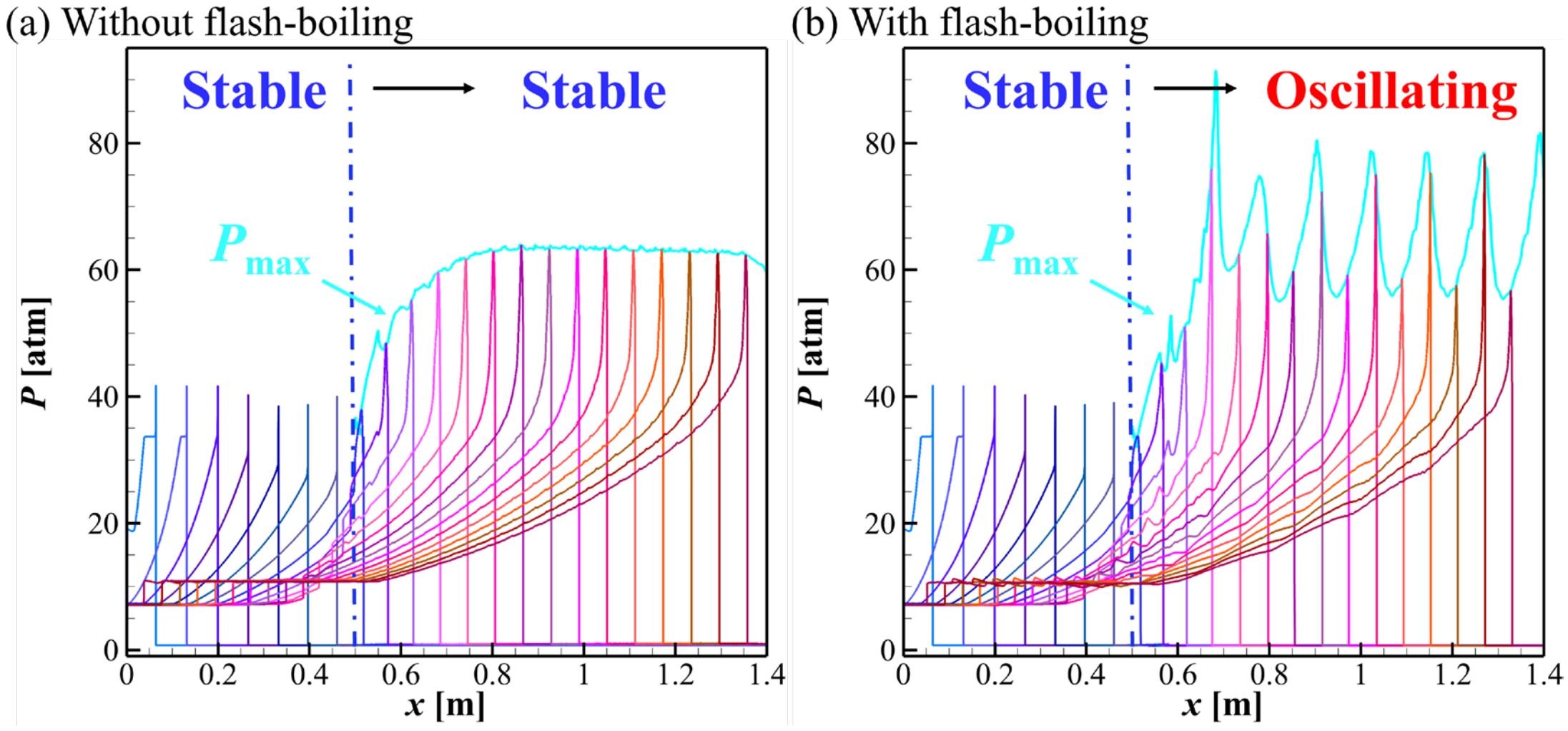

Fig. 3 Pressure evolution of the liquid-ammonia detonation under the reference condition ($T_{O_2}$ = 300 K, $D_d$ = 20 μm). The dashed line marks the end of the gaseous detonation zone, and the cyan curve traces the peak-pressure envelope, $P_{max}$, along the propagation path. (a) Without flash boiling, $P_{max}$ rises to a steady plateau after the front enters the droplet-laden region. (b) With flash boiling, $P_{max}$ develops pronounced periodic fluctuations, indicating oscillating propagation.

#### *3.1.2 Thermochemistry and chemical-kinetics interpretation*

To identify the thermochemical origin of the flash-boiling-induced oscillation, we compared the instantaneous wave structure with the composition of the reaction zone. Figure 4 shows the coupled gas-dynamic and dispersed-phase fields; Figure 5 summarises the dominant ammonia-oxidation pathway under the present post-shock conditions; and Figure 6 shows the mass fractions of diagnostic species $H_2NO$, OH, and $H_2O$. These species mark three successive stages of the reaction sequence. OH initiates $NH_3$ consumption through H abstraction, $H_2NO$ traces the pressure-favoured $NH_2$ oxidation channel, and $H_2O$ records the product-forming heat-release stage. Thus, the three fields provide a compact chemical measure of how flash boiling displaces, broadens, or weakens the reaction front relative to the leading shock.

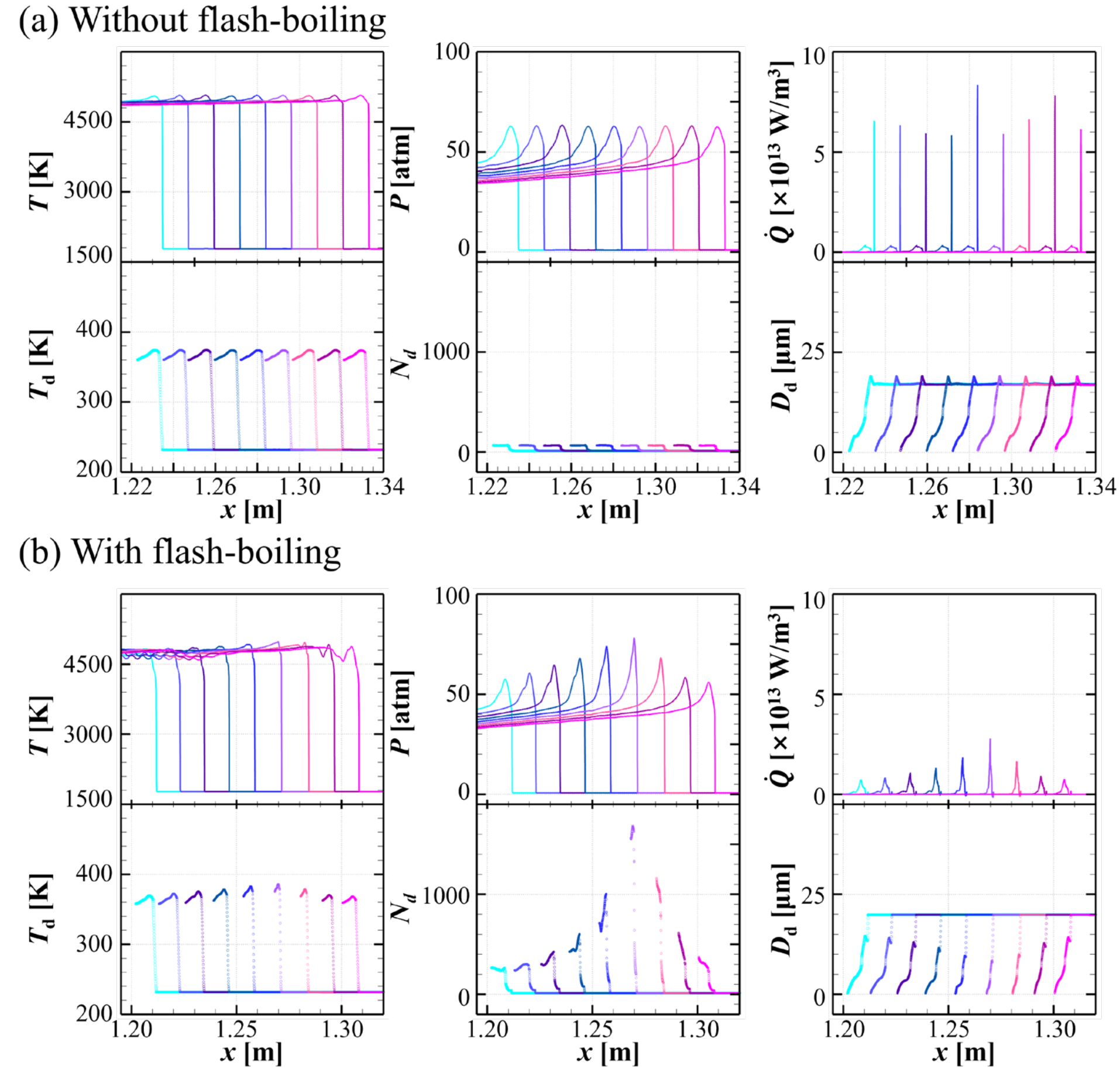

Fig. 4 Instantaneous detonation structure under the reference condition ($T_{O_2} = 300$ K, $D_d = 20$ μm). Fields: pressure, gas temperature, heat-release rate, droplet temperature, droplet number in a parcel, and droplet diameter. (a) Without flash boiling: uniform shock strength, compact and stationary heat-release peak. (b) With flash boiling: oscillating shock strength, broadened and migrating heat-release zone.

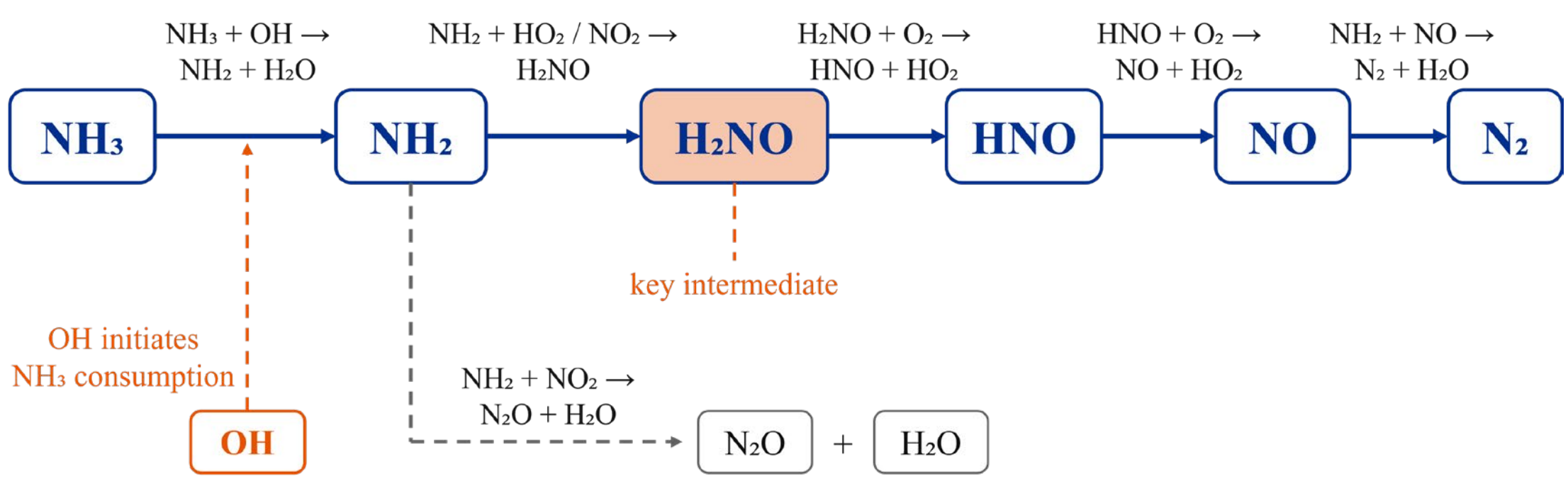

Fig. 5 Primary reaction pathway of ammonia oxidation under the present detonation-relevant conditions. Solid blue arrows trace the principal channel $NH_3 \rightarrow NH_2 \rightarrow H_2NO \rightarrow HNO \rightarrow NO \rightarrow N_2$, closed by the DeNOx step $NH_2 + NO \rightarrow N_2 + H_2O$. The orange dashed line indicates initiation by OH through (R1). The grey dashed branch through $NH_2 + NO_2 \rightarrow N_2O + H_2O$ contributes minor $N_2O$ but does not alter the main route. $H_2NO$, highlighted in salmon, is the diagnostic intermediate of the high-pressure $NH_2 + HO_2$ channel and provides the most sensitive marker of perturbations to the local $HO_2$ pool.

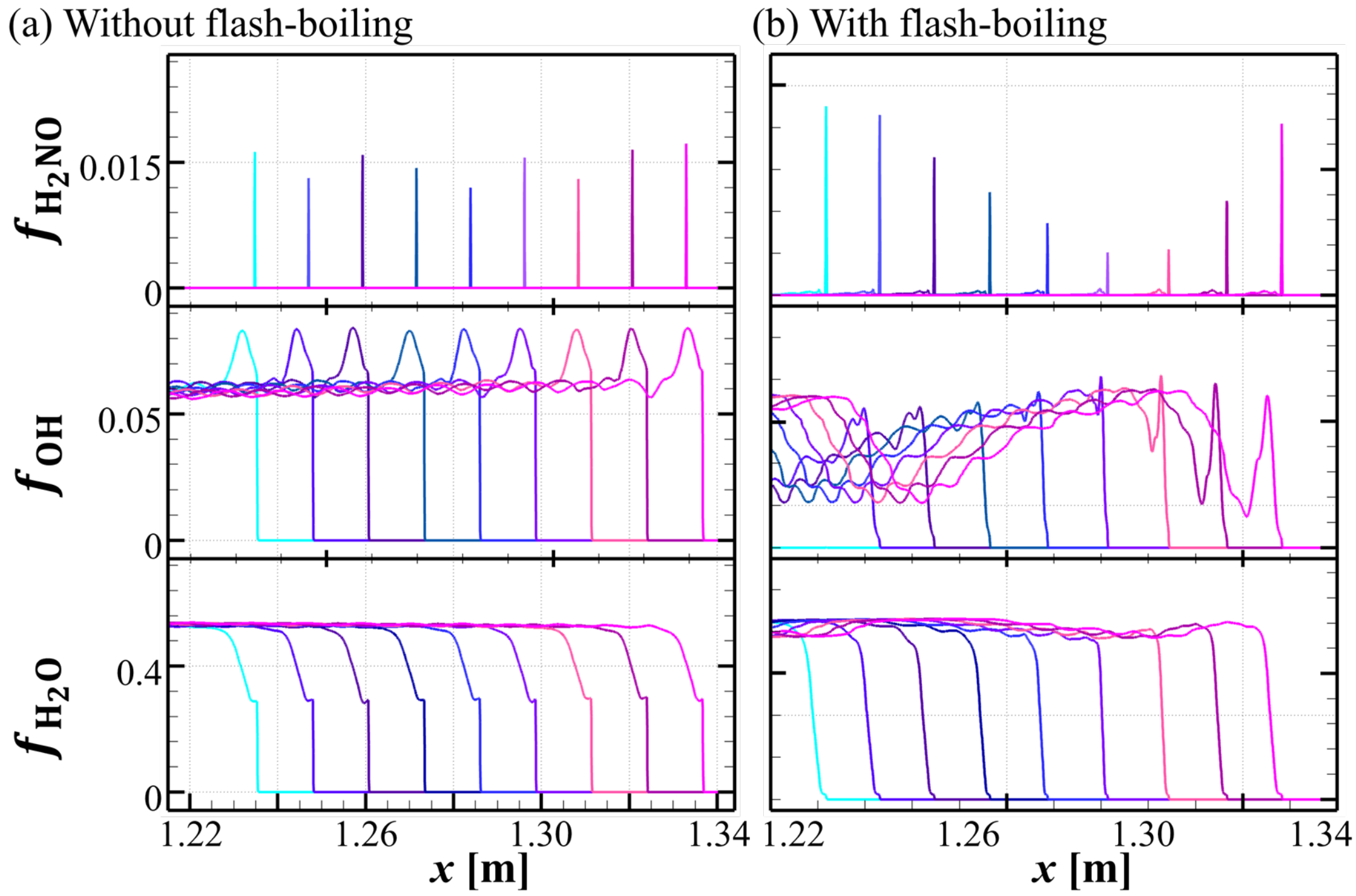

Fig. 6 Reaction-zone chemistry of the liquid-ammonia detonation under the reference condition $(T_{O_2} = 300\ \mathrm{K}, D_d = 20\ \mu\mathrm{m})$, showing profiles of $H_2NO$, OH, and $H_2O$. (a) Without flash boiling: invariant radical layers, steady product formation locked to the shock. (b) With flash boiling: oscillatory radical peaks, variable product-onset distance, periodic detachment of the reaction front.

Under the present detonation conditions, fuel consumption begins primarily through the H abstraction reaction, $NH_3 + OH \rightarrow NH_2 + H_2O$ (R1). $NH_2$ has several routes whose branching ratios are pressure dependent. In the present high-$HO_2$ environment, the channels $NH_2 + HO_2 \rightarrow H_2NO + OH$ (R2) and $NH_2 + NO_2 \rightarrow H_2NO + NO$ (R3) dominate over the low-pressure NH and NNH branches. The $H_2NO$ produced by (R2)–(R3) is therefore the diagnostic intermediate of high-pressure $NH_3$ oxidation, and it is rapidly oxidised through $H_2NO + O_2 \rightarrow HNO + HO_2$ (R4), which regenerates $HO_2$ and feeds (R2) in a self-amplifying chain. The nitrosyl HNO is converted to NO via $HNO + O_2 \rightarrow NO + HO_2$ (R5), after which the bulk of the heat release proceeds through the DeNOx step, $NH_2 + NO \rightarrow N_2 + H_2O$ (R6); a side branch $NH_2 + NO_2 \rightarrow N_2O + H_2O$ (R7) contributes minor amounts of $N_2O$ without altering the principal route.

Two consequences of this pathway guide the species selection in Figure 6. First, the global fuel-consumption rate is gated by OH through (R1), so any thermal perturbation that depresses OH in the immediate post-shock region broadens the induction zone and weakens the shock–reaction coupling. Second, $H_2NO$ is the most sensitive marker of the high-pressure $NH_2 + HO_2$ channel: a disturbance to the local $HO_2$ pool registers first as a depression of $H_2NO$, and only subsequently as a delay in $H_2O$ formation. The triplet (OH, $H_2NO$, $H_2O$) therefore resolves initiation, high-pressure intermediate formation, and terminal heat release, respectively. It provides the chemical basis for interpreting the oscillatory displacement of the reaction zone observed in figures 4 and 6.

Without flash boiling, the leading shock maintains uniform strength, evidenced by the constant amplitude of the von Neumann pressure spikes ($P \approx 65$ atm) and post-shock temperatures. This hydrodynamic stability stems from a coherent coupling between droplet breakup, evaporation, and the reaction zone structure. As observed in figures 4(a) and 6(a), droplets undergo steady disintegration, heating, and vaporisation behind the shock, providing a continuous supply of gaseous $NH_3$. The OH radical profiles exhibit a sharp post-shock rise followed by a gradual decay, mapping the trajectory of $NH_3$ consumption and the initiation of primary reaction pathways; crucially, these profiles remain invariant across different time instants. Since OH drives the initial H-abstraction from $NH_3$, its sustained high concentration ensures rapid, uniform fuel conversion to $H_2O$, effectively locking the reaction front to the shock wave. Although minor fluctuations appear in the $H_2NO$ intermediate and heat release rate profiles, the generation of the final product, $H_2O$, remains steady.

A qualitatively different structure is obtained for the case with flash boiling. In figure 4(b), the larger $N_d$ coincides with a more pronounced temperature deficit immediately behind the shock, which is caused by the intensified latent-heat extraction during the rapid phase change. In addition, the heat-release peak no longer remains fixed relative to the wave front. Instead, its magnitude, sharpness, and wave-relative position vary substantially over the cycle. Figure 6(b) shows the chemical counterpart of these unsteady features—the cycle-to-cycle modulation of shock strength, heat-release magnitude, and heat-release position. The $H_2NO$ peak becomes strongly oscillatory, the OH layer is broadened and distorted, and the onset of $H_2O$ formation no longer occurs at a unique distance behind the shock. These indicate that flash boiling affects the combustion chemistry by altering the local thermal state and the composition field in the near-shock region. The early high-temperature $NH_3 \rightarrow NH_2 \rightarrow H_2NO$ oxidation sequence and the subsequent product-forming heat-release stage are advected to varying distances behind the leading front: their separation contracts during phases of stronger shock support and dilates during phases of weakened support, as evident from the migration of the heat-release peak in figure 4(b) and from the wandering OH/$H_2NO$ layers in figure 6(b).

Based on the above analysis, the oscillation is sustained by a periodically varying near-shock thermal penalty, which is transmitted to the chemistry through both temperature and species evolution. During the flash boiling, the larger latent-heat burden depresses the immediate post-shock temperature, modifies the local radical/intermediate structure, weakens the compact high-temperature heat release close to the shock, and shifts the main product-forming region farther downstream. During phases in which the latent-heat sink locally weakens—when evaporation length extends and the droplet number density immediately behind the shock falls, the near-shock reactive structure becomes more compact again, the $H_2NO$ and OH layers recover a sharper localisation, and $H_2O$ formation moves closer to the front. The oscillation therefore arises because flash boiling periodically reorganises the intensity, timing and spatial distribution of ammonia oxidation and heat release relative to the shock, thereby driving repeated strengthening and weakening of shock–reaction coupling.

#### *3.1.3 Temporal signatures and phase-space characterisation*

Figure 7 characterises the fully developed detonation dynamics over $t$ = 0.35 – 0.5 ms by combining time histories with a phase-space representation. The evaporation length, $L_v$, is defined in the shock-attached frame as the distance from the leading shock to the location at which the liquid droplets are completely depleted. In the case without flash boiling, shown in Figure 7(a), the trajectory in the $(P_{\max}, L_{\mathrm{v}})$ plane collapses into a compact cluster. The corresponding time histories show only weak variations: $P_{\max}$ remains close to 63 atm, and $L_{\mathrm{v}}$ stays near 12 mm throughout the sampled interval. This compact phase portrait confirms a quasi-steady propagation state, in which both the shock strength and the spatial extent of the evaporation zone remain nearly invariant after the initial transmission transient.

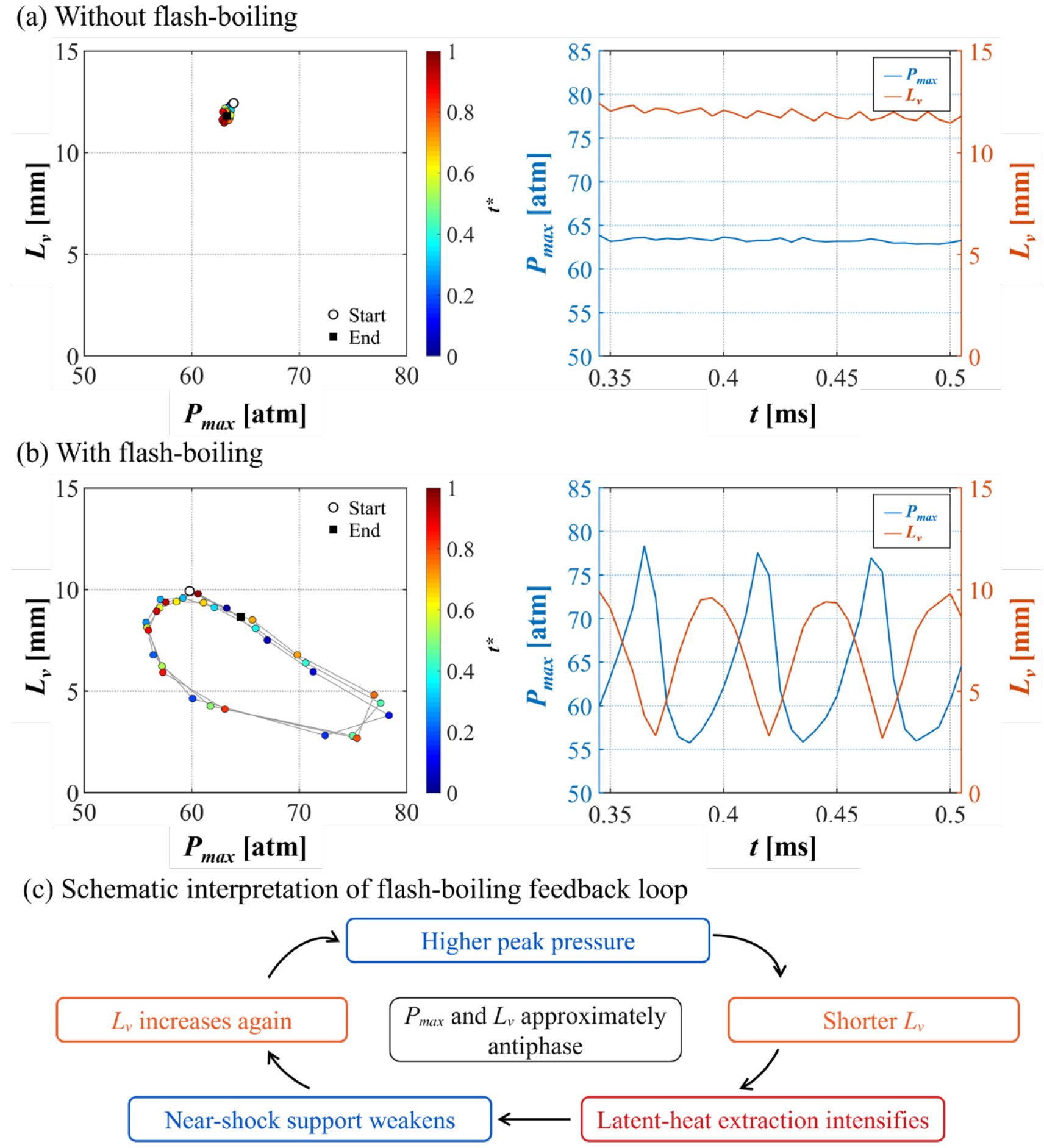

Fig. 7 Temporal and phase-space signatures of the liquid-ammonia detonation dynamics over the fully developed window $t$ = 0.35–0.5 ms. Left columns: trajectory in the $(P_{\max}, L_{\mathrm{v}})$ plane, with markers coloured by normalized time $t^*$ and the start/end states indicated. Right columns: the corresponding time histories of $P_{\max}$ and $L_{\mathrm{v}}$. (a) Without flash boiling, the trajectory collapses into a compact cluster and both signals are nearly stationary, consistent with quasi-steady propagation. (b) With flash boiling, the trajectory forms a closed orbit and $P_{\max}$ and $L_{\mathrm{v}}$ vary approximately in antiphase, consistent with a self-sustained oscillatory mode. (c) Schematic of the feedback cycle: a stronger shock contracts the evaporation zone, which concentrates the latent-heat sink immediately behind the front, weakens the near-shock thermochemical support, and forces the shock to relax; droplet depletion then slows, $L_{\mathrm{v}}$ re-expands, and the cycle repeats.

Flash boiling fundamentally changes this dynamical signature. As shown in Figure 7(b), the trajectory does not collapse to a fixed point but forms a distinct closed orbit. The time traces reveal an approximately antiphase relation between $P_{\max}$ and $L_{\mathrm{v}}$: pressure maxima coincide with a contracted evaporation zone, whereas pressure minima occur when the evaporation zone has re-expanded. The closure of the orbit indicates that the wave has entered a self-sustained oscillatory state, rather than undergoing a transient adjustment.

Figure 7(c) summarises the feedback loop responsible for this limit-cycle-like response. A momentarily stronger shock raises the post-shock temperature and increases the droplet–gas slip velocity. Both effects accelerate droplet breakup, heating, and vaporisation, so complete droplet depletion occurs closer to the front and $L_{\mathrm{v}}$ contracts. For a fixed incoming droplet loading, the latent-energy demand associated with liquid depletion is set primarily by the amount of liquid carried into the shock. When $L_{\mathrm{v}}$ becomes shorter, this endothermic demand is imposed over a shorter shock-attached distance; the latent-heat removal per unit length in the immediate post-shock region therefore increases. This concentrated cooling competes directly with chemical heat release, depresses the near-shock temperature, broadens the induction and reaction zones, and weakens the thermochemical support that sustains the leading shock. The weakened shock then reduces the thermal and aerodynamic forcing on the droplets. Droplet depletion slows, $L_{\mathrm{v}}$ increases again, and the latent-heat sink spreads over a longer region behind the front. Near-shock cooling correspondingly weakens, allowing the radical pool and product-forming heat release to recover a more compact structure. The shock is then re-strengthened, and the cycle repeats. The phase portrait therefore provides direct evidence that the flash-boiling-induced oscillation is governed by a feedback between heat release and evaporation localisation.

#### *3.1.4 Parametric dependence of oscillating detonation*

To assess how the propagation response depends on the ambient temperature and droplet diameter, Figure 8 compares the time histories of the peak shock pressure, $P_{\max}$, for four representative cases, both without and with flash boiling. The dominant oscillation frequency $f$ and the minimum of evaporation length $\min(L_v)$ extracted from each case are indicated directly in the panels.

For the cases without flash boiling, the system exhibits a clear bifurcation in stability governed by the evaporation timescale. The baseline condition ($T_{O_2} = 300$ K, $D_d = 20$μm) sustains a stable detonation, where the evaporation rate is sufficient to stably couple the reaction zone with the shock front. However, perturbing this state by either increasing the droplet diameter to 40 μm or reducing the ambient temperature to 220 K triggers the onset of instability. Physically, both perturbations extend the characteristic evaporation time—one via a reduced surface-area-to-volume ratio and the other via a diminished thermal driving force. This retardation decouples the energy release from the shock, initiating oscillatory behaviour. The oscillation frequency exhibits a strong sensitivity to droplet size in this regime; at $T_{O_2} = 220$ K, doubling the diameter from 20 μm to 40 μm approximately doubles the frequency from 15.15 kHz to 30.17 kHz. This trend correlates with the contraction of $\min(L_v)$, which decreases from 4.62 mm to 3.39 mm, indicating that larger droplets, paradoxically, trigger more compact and violent heat release excursions during the cycle's compression phase.

The cases with flash boiling show an oscillating propagation mode across the entire investigated parameter space. Here, the rapid, endothermic phase change inherently destabilises the detonation propagation regardless of the specific droplet diameter or thermal conditions. However, the oscillation characteristics vary significantly. Increasing the droplet diameter consistently elevates the oscillation frequency (e.g., from 19.20 kHz to 21.29 kHz at 300 K). However, lowering the ambient temperature in the flash-boiling regime results in a marked reduction in frequency (e.g., dropping from 19.20 kHz to 10.40 kHz for $D_d$ = 20 μm). This distinct behaviour highlights the complex, nonlinear coupling between the vaporisation intensity, the heat release rate, the induction delay, and the acoustic transit time within the reaction zone. Comparisons of $L_v$ further reveal that the flash-boiling model generally yields shorter minimum evaporation lengths than the conventional model (e.g., 2.68 mm vs. 11.69 mm at 300 K/20 μm), confirming that the instability is driven by rapid, compact vaporisation that is thermally quenching.

Fig. 8. Dependence of the liquid-ammonia detonation on the ambient temperature and droplet diameter. Top row: cases without flash boiling. Bottom row: cases with flash boiling. Each panel shows the time history of the peak shock pressure, $P_{\max}$, for one representative case. The dominant oscillation frequency, $f$, and the minimum evaporation length, $\min(L_v)$, extracted from each case are indicated in the panels.

*3.2 Delay-differential-equation stability model*

*3.2.1 Theoretical considerations*

The oscillation of the leading shock is caused by two delayed thermal processes sharing a single shock front. Droplet evaporation supplies fuel vapour that sustains the downstream chemical reaction, and the resulting heat release reinforces the shock; latent-heat absorption by the same droplets, on the other hand, cools the post-shock gas and weakens the front. Since reaction also modifies the thermal state that drives evaporation, the two processes are coupled in both directions with finite response delays. A finite-frequency shock response associated with this coupled two-phase mechanism is selected most clearly when the delayed couplings have comparable response times and retain a non-zero phase difference; if either process becomes much faster than the other, this two-process phase relation collapses. The remainder of this theoretical section places this mechanism on a quantitative footing through a DDE model that retains only a shock front, two delayed thermal processes and a closed dynamical system relating them. The analysis takes the form of a standard normal-mode stability problem for this DDE system, in the spirit of the time-lag combustion-instability framework pioneered by Tsien [73] and Crocco and Cheng [74]. In the asymptotic limit of a purely gaseous inflow the formulation collapses to the canonical single-phase stability models of Whitham [75], and Wang, Xiang, and Jiang [76].

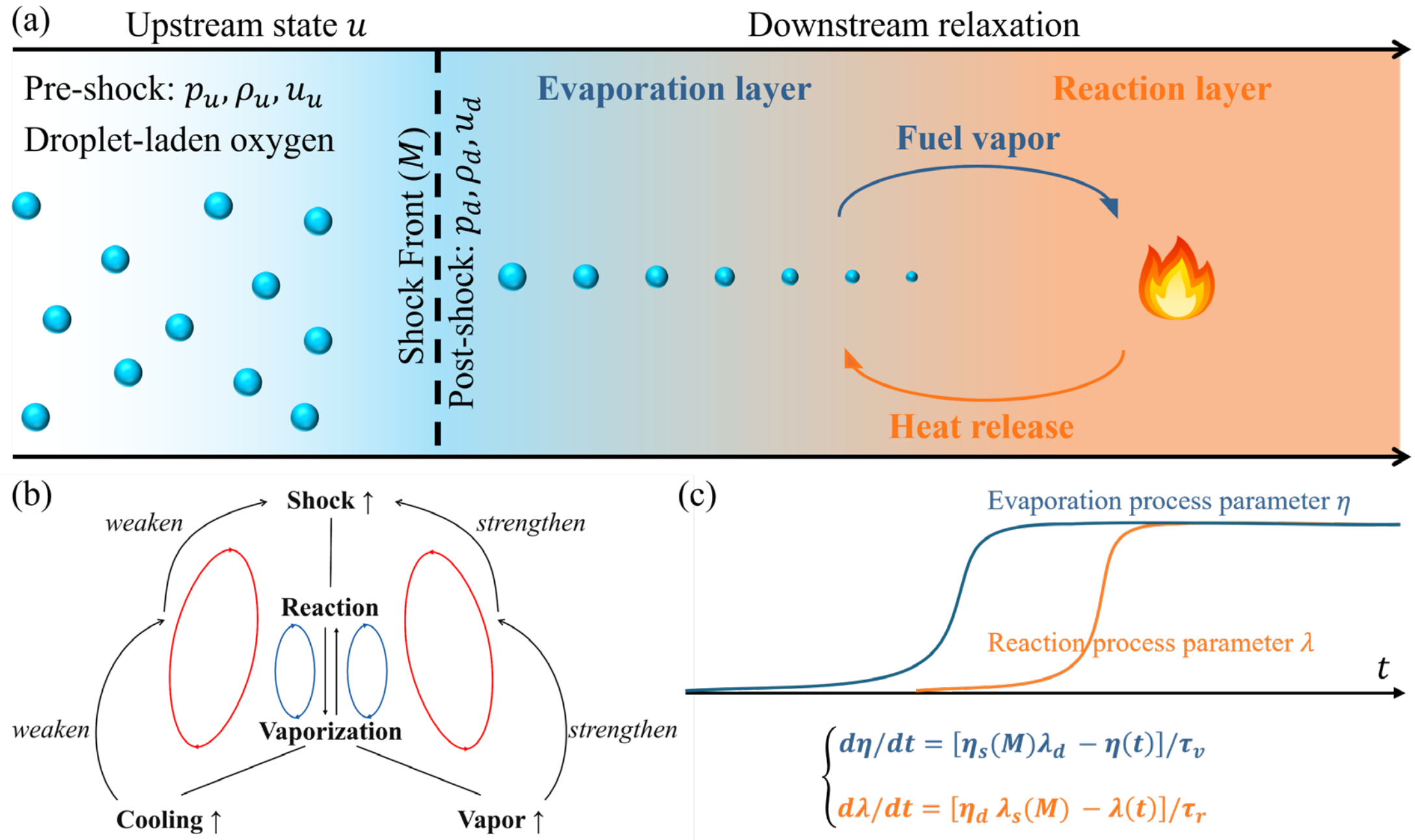

Fig. 9 Physical structure underlying the reduced DDE stability model. The upstream state consists of quiescent droplet-laden oxygen ahead of the leading shock. The shock, with instantaneous Mach number $M$, produces the post-shock state, within which an evaporation layer, represented by progress variable $\eta$, supplies fuel vapor to a downstream reaction layer, represented by progress variable $\lambda$. The delayed relaxation equations for $\eta$and $\lambda$are shown below the schematic.

The one-dimensional structure relative to the front is sketched schematically in figure 9(a):

the upstream state $(p_u, \rho_u, u_u)$ of the droplet-laden oxidiser is compressed at the shock to $(p_d, \rho_d, u_d)$, droplets vaporise across the evaporation layer, and the reaction layer further downstream completes the heat-release cycle. The schematic of the coupled feedback is shown in figure 9(b). Heat release strengthens the shock and raises the post-shock thermal state; through the delayed thermal response, a stronger reaction promotes vaporisation, whereas a weaker reaction suppresses it. Vaporisation then feeds back through two channels. The latent-heat sink draws heat from the post-shock region ("cooling") and tends to weaken the front, whereas the added fuel vapour ("vapour") promotes downstream reaction and can strengthen the heat-release feedback. The evaporation pathway therefore has no single fixed sign; its net effect is set by the delayed balance between cooling, vapour supply and reaction. The delay on the reaction response is set by the finite time required for droplet heating and evaporation to supply fuel vapour to the reaction layer, while the delay on the evaporation response reflects the finite time required for reaction heat release to modify the thermal drive for vaporisation. These two delays are intrinsic to the relaxation-zone coupling and are not free fitting parameters.

To make the picture quantitative we follow the evaporation and reaction progress along a fluid element behind the shock through two scalar variables $\eta(t)$ and $\lambda(t)$, which rise from zero in the unprocessed droplet-laden stream to unity once the corresponding process has run to completion. Their typical histories are sketched in figure 9(c); the offset between the two profiles is a schematic representation of the finite phase difference upon which the analysis rests. The relaxation law displayed in the same panel is introduced formally below.

*3.2.2 Mathematical formulation*

*3.2.2.1 Shock-response constraint*

The theory rests upon the compatibility condition carried along the $C_+$ characteristic of the shock, which governs the evolution of the flow field immediately behind the front. We use $Q_r$ and $Q_v$ from the outset for the reduced reaction and vaporisation heat densities, obtained by scaling the corresponding effective energy densities by the reference energy density $\rho_u a_u^2/(\gamma - 1)$. Equivalently, the thermodynamic factor $(\gamma - 1)$ multiplying the heat source has already been absorbed into $Q_r$ and $Q_v$. With the pressure and momentum terms scaled by $\rho_u a_u^2$, the compatibility condition is

$$\frac{1}{\rho_u a_u^2}(\dot{p} + \rho a \, \dot{u}) - \left(Q_r \dot{\lambda} - Q_v \dot{\eta}\right) = 0, \tag{3.1}$$

where the over dot denotes differentiation along this $C_+$ characteristic. Here $p$, $\rho$, $u$ and $a$ denote pressure, density, velocity and local sound speed, respectively, and the rates $\dot{\lambda}$ and $\dot{\eta}$ are associated with the chemical and evaporation progress variables introduced in § 3.2.1. The source term in (3.1) thus expresses, in equation form, the competition between exothermic chemical heat release and endothermic droplet vaporisation sketched in Figure 9. Differentiating the Rankine–Hugoniot jump conditions for a shock propagating at instantaneous Mach number $M(t)$ provides $\mathrm{d}p_d = 4\gamma p_u M \,\mathrm{d}M/(\gamma + 1)$, $\mathrm{d}u_d = 2a_u(1 + M^{-2})\,\mathrm{d}M/(\gamma + 1)$, and $\rho_d a_d = \rho_u a_u M/\mu(M)$, in which subscripts $u$ and $d$ refer, throughout this paper, to the quiescent upstream and compressed downstream states. Substituting the post-shock relations into (3.1) casts the shock-evolution equation in the shock-response form

$$f(M)\,\dot{M} = Q_r\,\dot{\lambda} - Q_v\,\dot{\eta}, \tag{3.2}$$

where

$$f(M)=\frac{2M}{\gamma+1}\left[2+\frac{1}{\mu(M)}\left(1+\frac{1}{M^2}\right)\right],\mu(M)=\left[\frac{(\gamma-1)M^2+2}{2\gamma M^2-(\gamma-1)}\right]^{\frac{1}{2}} \tag{3.3}$$

$\mu(M)$ is the post-shock Mach-number factor, of order unity (close to one third) for the strong shocks of present interest, and $f(M)$ is the resulting dimensionless shock-response coefficient obtained from the scaled pressure and momentum terms. In Equation (3.2) the first right-hand term represents the accelerating effect of chemical heat release on the shock and the second the latent-heat sink.

*3.2.2.2 Relaxation-model closure*

Let $\eta_s(M)$ and $\lambda_s(M)$ denote the corresponding quantities along the non-oscillatory two-phase branch, regarded as differentiable functions of $M$ in a neighbourhood of a chosen reference Mach number $M_0$. The delayed responses of evaporation and reaction to changes in shock strength are characterised by two response delays $\tau_{v,d}$ and $\tau_{r,d}$, in terms of which we introduce the delayed response factors

$$\lambda_d(t)=\frac{\lambda(t-\tau_{r,d})}{\lambda_s[M(t-\tau_{r,d})]},\qquad \eta_d(t)=\frac{\eta(t-\tau_{v,d})}{\eta_s[M(t-\tau_{v,d})]}, \tag{3.4}$$

each of which is dimensionless and identically unity on the steady branch.

The closure of the system, anticipated in Figure 9(c), is supplied by the assumption that the evaporation and reaction progress variables relax, on their own characteristic times $\tau_v$ and $\tau_r$, towards branch values modulated by the delayed response factor of the opposite loop:

$$\dot{\eta}=\frac{\lambda_d\,\eta_s(M)-\eta}{\tau_v},\qquad \dot{\lambda}=\frac{\eta_d\,\lambda_s(M)-\lambda}{\tau_r}. \tag{3.5}$$

The numerators in (3.5) are dimensionless progress-variable differences, so $\tau_v$ and $\tau_r$ are the relaxation times that set the dimensions of $\dot{\eta}$ and $\dot{\lambda}$. The closure (3.5) is the mathematical embedding of the bidirectional coupling introduced above: the delayed reaction response modulates the evaporation target through $\lambda_d$, while the delayed evaporation response modulates the reaction target through $\eta_d$. Thus, a stronger reaction enhances the thermal drive for evaporation, and increased vaporisation supplies fuel vapour that can strengthen the reaction, with both influences entering only after their respective response delays. The multiplicative form $\lambda_d\,\eta_s(M)$ is the simplest closure that preserves the steady branch as a fixed point of (3.5); the delay parameters $\tau_{r,d}$ and $\tau_{v,d}$ retain the geometric interpretation described in Section 3.2.1.

*3.2.2.3 Non-dimensionalisation*

The pair (3.5), together with the shock-response constraint (3.3) and the delay relations (3.4), forms a DDE system. We non-dimensionalise time by $\tau_r$, retain $t$ as the (dimensionless) time variable to avoid clumsy notations, reinterpret the overdot as differentiation with respect to this time, and introduce the Damköhler number

$$\mathrm{Da}=\frac{\tau_v}{\tau_r}, \tag{3.6}$$

the ratio of the evaporation time to the reaction time, together with the two delay parameters $\delta_r = \tau_{r,d}/\tau_r$ and $\delta_v = \tau_{v,d}/\tau_v$. Large Da corresponds to slow evaporation relative to the reaction (the chemical-equilibrium regime in which the reaction adjusts almost instantaneously to the slowly evolving vapour supply), and small Da to fast evaporation (the chemically frozen regime in which the latent-heat sink is slaved to the instantaneous post-shock state). The closed system then takes the form

$$f(M)\,\dot{M} = Q_r\,\dot{\lambda} - Q_v\,\dot{\eta}, \tag{3.7}$$

$$\dot{\eta} = \frac{\lambda_d\,\eta_s(M) - \eta}{\mathrm{Da}}, \qquad \dot{\lambda} = \eta_d\,\lambda_s(M) - \lambda, \tag{3.8}$$

with the delay factors of (3.4) becoming $\lambda_d(t) = \lambda(t-\delta_r)/\lambda_s[M(t-\delta_r)]$ and $\eta_d(t) = \eta(t-\mathrm{Da}\delta_v)/\eta_s[M(t-\mathrm{Da}\delta_v)]$. In (3.7)–(3.8), the reaction relaxation has been absorbed into the unit time scale, while the evaporation relaxation time is now Da, and the same parameter positions the evaporation delay $\mathrm{Da}\delta_v$ relative to the reaction delay $\delta_r$ in the time-shifted arguments of the delay factors. Both $\delta_r$ and $\delta_v$ remain order-unity geometric factors, intrinsic to the relaxation zone, so that the relative phase between the two delays in figure 9(b) is controlled almost entirely by Da: the regimes $\mathrm{Da} \ll 1$ and $\mathrm{Da} \gg 1$ collapse this phase, while $\mathrm{Da} = O(1)$ preserves it.

*3.2.2.4 Perturbation about non-oscillating states*

Let $(M_0, \eta_0, \lambda_0)$ be a point on the non-oscillatory branch and write

$$\begin{aligned} M(t) &= M_0 + M_1(t),\\ \eta(t) &= \eta_0 + \eta_1(t) = \eta_s(M_0) + \eta_1(t),\\ \lambda(t) &= \lambda_0 + \lambda_1(t) = \lambda_s(M_0) + \lambda_1(t), \end{aligned} \tag{3.9}$$

where the perturbed quantities are denoted with subscript "1", and $|M_1(t)|\& \ll M_0$. Setting $k_\eta = \mathrm{d}\eta_s/\mathrm{d}M|_{M_0}$ and $k_\lambda = \mathrm{d}\lambda_s/\mathrm{d}M|_{M_0}$, the linearised forms of the delay factors are

$$\lambda_d = 1 + \frac{\lambda_1(t-\delta_r) - k_\lambda M_1(t-\delta_r)}{\lambda_0}, \quad \eta_d = 1 + \frac{\eta_1(t-\mathrm{Da}\delta_v) - k_\eta M_1(t-\mathrm{Da}\delta_v)}{\eta_0}. \tag{3.10}$$

Substituting (3.9) and (3.10) into (3.7)–(3.8) and retaining the perturbed quantities, we obtain

$$f(M_0)\,\dot{M}_1 = Q_r\,\dot{\lambda}_1 - Q_v\,\dot{\eta}_1, \tag{3.11}$$

$$\dot{\eta}_1 = \frac{1}{\mathrm{Da}}\left\{-\eta_1 + k_\eta M_1 + \frac{\eta_0}{\lambda_0}\left[\lambda_1(t-\delta_r) - k_\lambda M_1(t-\delta_r)\right]\right\}, \tag{3.12}$$

$$\dot{\lambda}_1 = -\lambda_1 + k_\lambda M_1 + \frac{\lambda_0}{\eta_0}\left[\eta_1(t-\mathrm{Da}\delta_v) - k_\eta M_1(t-\mathrm{Da}\delta_v)\right]. \tag{3.13}$$

The leading $-\eta_1/\mathrm{Da}$ in (3.12) and $-\lambda_1$ in (3.13) are the natural-decay contributions inherited from the relaxation closure, while the remaining terms couple the local branch slopes to the delayed cross-loop forcing. Equation (3.11) integrates immediately. The integration constant is the projection of the initial perturbation onto the tangent of the steady branch at $(M_0, \eta_0, \lambda_0)$; this corresponds to a relabelling of the reference state rather than a stability mode, so we set it to zero and obtain

$$M_1 = \frac{Q_r\,\lambda_1 - Q_v\,\eta_1}{f(M_0)}, \tag{3.14}$$

which expresses $M_1$ algebraically in terms of the two progress perturbations and may be used to eliminate $M_1$ from (3.12)–(3.13).

*3.2.2.5 Characteristic equation*

We seek modal solutions

$$(M_1, \eta_1, \lambda_1)(t) = \left(\hat{M}, \hat{\eta}, \hat{\lambda}\right) \mathrm{e}^{st}, \tag{3.15}$$

so that $\eta_1(t-\delta_r) = \hat{\eta}\exp[s(t-\delta_r)]$ and $\lambda_1(t - Da\delta_v) = \hat{\lambda}\exp[s(t - Da\delta_v)]$. The eigenvalue $s$ is in general complex; its real part determines the temporal growth or damping rate of the disturbance, and its imaginary part, when non-zero, the angular frequency. A positive real part with non-zero imaginary part would signal a linearly self-excited oscillation, whereas a negative real part with non-zero imaginary part represents a damped finite-frequency response. On eliminating $M_1$ via (3.14), the system (3.12)–(3.13) becomes a homogeneous algebraic system in $(\hat{\eta}, \hat{\lambda})$. We introduce the four shock-normalised parameters

$$q_r = \frac{Q_r\, k_\lambda}{f(M_0)}, \qquad q_v = \frac{Q_v\, k_\eta}{f(M_0)}, \qquad \chi = \frac{Q_v}{Q_r}, \qquad \kappa = \frac{\eta_0}{\lambda_0}, \tag{3.16}$$

in which $q_r$ and $q_v$ measure the strength of each branch response in units of the shock acceleration, and the heat-and-progress ratio $\chi\kappa$ collects the relative reference levels of the two channels. After dividing through by a non-vanishing constant, the requirement of a non-trivial modal amplitude reduces to the single characteristic equation $\mathcal{D}(s) = 0$, which on grouping by the underlying delay channel takes the form

$$\begin{aligned}\mathcal{D}(s) = \; & \left[\mathrm{Da}\, s^2 + \{1 + q_v + \mathrm{Da}(1-q_r)\}\, s + (1 - q_r + q_v)\right] \\ & - \chi\kappa q_r\, s\, \mathrm{e}^{-s\delta_r} + \mathrm{Da}\, q_v\, \chi^{-1}\kappa^{-1} s\, \mathrm{e}^{-s\,\mathrm{Da}\delta_v} - (1 - q_r + q_v)\, \mathrm{e}^{-s(\delta_r + \mathrm{Da}\delta_v)} = 0.\end{aligned} \tag{3.17}$$

The four contributions in (3.17) carry distinct physical meaning: the bracketed polynomial is the instantaneous response of the closed system; the term in $\exp(-s\delta_r)$ is the contribution of the reaction-side delay alone; the term in $\exp(-s\,\mathrm{Da}\delta_v)$ is the contribution of the evaporation-side delay alone; and the term in $\exp[-s(\delta_r + \mathrm{Da}\delta_v)]$ is the cross-coupling of the two delays. Equation (3.17) contains the neutral root $s = 0$ associated with the integration constant of (3.11); this root is excluded below when extracting the stability modes. The exponentials are retained in full because the oscillatory modes of physical interest possess $s\delta_r = O(1)$ or $s\mathrm{Da}\delta_v = O(1)$; a low-frequency Taylor expansion would suppress the very phase lags responsible for the oscillation and is therefore inappropriate as a stability model.

The combination $1 - q_r + q_v$ that appears in both the polynomial and the cross-coupling term plays a special role in what follows. At $1 - q_r + q_v = 0$, or equivalently $f(M_0) + Q_v k_\eta = Q_r k_\lambda$, the reaction-branch gain balances the combined shock response and evaporative branch loss. The asymptotic analysis below therefore amounts to an expansion about this neutral branch-balance curve, and the leading-order roots describe how the modal decay or growth rate changes as the system is detuned from this locus.

*3.2.2.6 Asymptotic limits and resonance window*

Two distinguished limits of $Da$ admit a tractable asymptotic analysis and bracket the response of the system; the algebra is given in Appendix A. In both limits, the leading non-trivial principal root on the branch-balance locus is real, so neither endpoint supports a finite-frequency coupled reaction–vaporisation resonance. Physically, in the fast-vaporisation limit $\mathrm{Da} \ll 1$ the latent-heat sink is slaved to the instantaneous post-shock state and no longer supplies an independent delayed leg of the feedback cycle; in the slow-vaporisation limit $Da \gg$ 1 the gas-phase reaction adjusts almost instantaneously to the vapour supplied by the slowly evolving droplets, so it is the reaction zone that ceases to form an independent delayed leg.

The distinguished regime is consequently $Da = O(1)$. In this range neither reaction nor evaporation can be eliminated as an instantaneous response, and the two delay factors $\exp(-s\delta_r)$ and $\exp(-s\,\mathrm{Da}\delta_v)$ in (3.17) carry comparable phase lags. This can be seen analytically by taking the representative slice $Da = 1$ and $\delta_r = \delta_v = \delta$. On the branch-balance locus $1 - q_r + q_v = 0$, the neutral mode $s = 0$ is removed and (3.17) reduces to the scalar delay equation,

$$s + 1 + B\,\mathrm{e}^{-s\delta} = 0. \tag{3.18}$$

Here $B = q_v/(\chi\kappa) - \chi\kappa q_r$ measures the net delayed feedback. Substituting $s = \mathrm{i}\Omega$ into (3.18) gives $1 + B\cos(\Omega\delta) = 0$ and $\Omega - B\sin(\Omega\delta) = 0$, so that, for $B > 1$,

$$s_{\pm} = \pm\mathrm{i}\sqrt{B^2 - 1}, \qquad \delta_0 = \frac{\arccos(-1/B)}{\sqrt{B^2 - 1}}. \tag{3.19}$$

Thus, the reduced model has an exact finite-frequency oscillatory pair in the comparable-time-scale regime. This result is the point needed here: the retained reaction–vaporisation phase difference can itself generate an oscillatory shock mode.

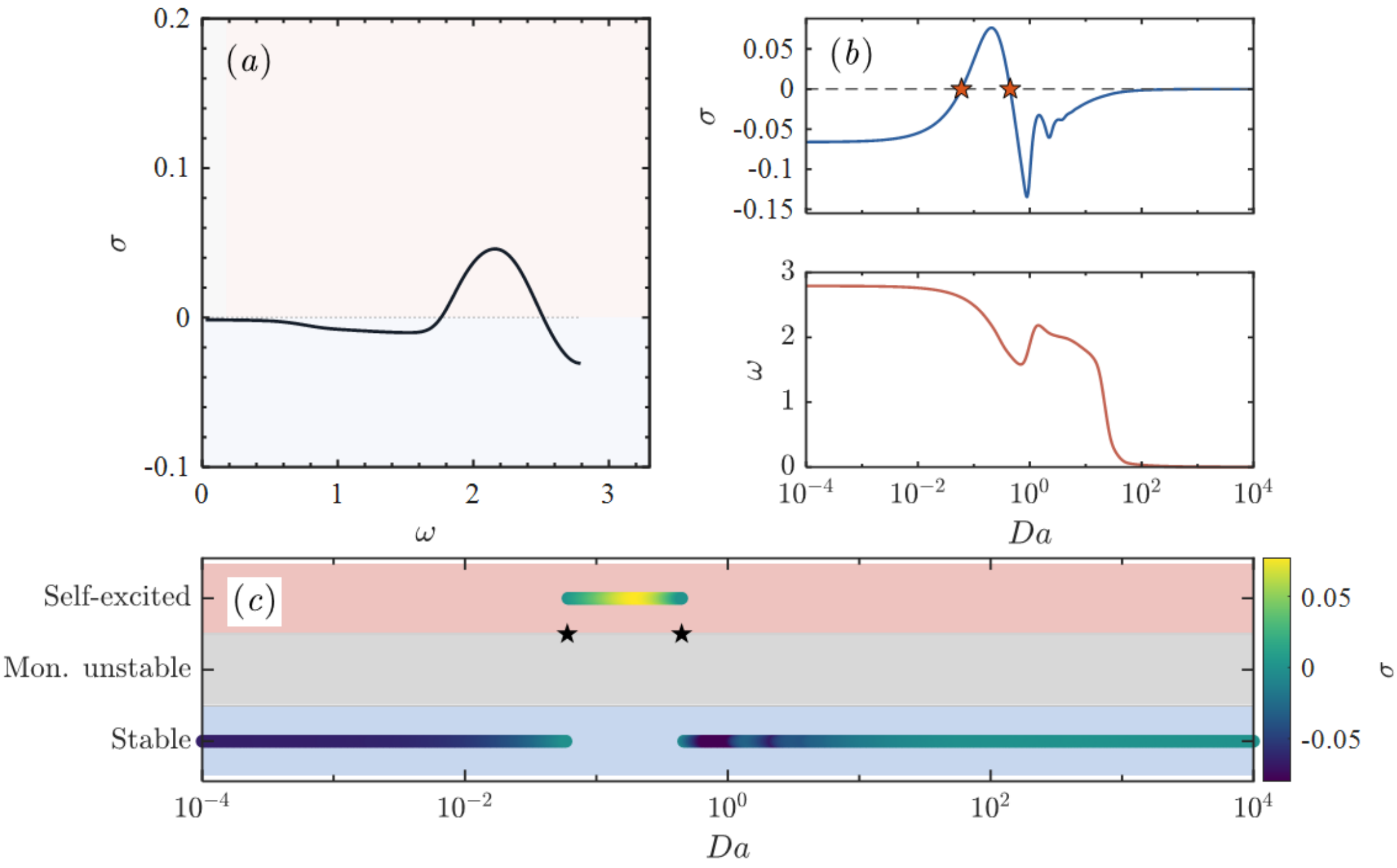

Fig. 10 DDE-predicted stability diagram of the dominant branch for the baseline parameters. (a) Parametric locus in the $(\omega, \sigma)$plane as $Da$ varies. (b) Growth rate, $\sigma$, and angular frequency, $\omega$, as functions of $Da$; stars mark finite-frequency neutral crossings. (c) Regime classification along $Da$. The dominant DDE branch is stable in both asymptotic limits and becomes self-excited only within an intermediate finite-frequency window.

### *3.2.3 Numerical analysis*

#### *3.2.3.1 Stability transition with $Da$*

We first solve the dimensionless DDE characteristic equation derived in Section 3.2.2.5 to determine how the dominant branch varies with $Da$. For a fixed operating condition, the upstream and equilibrium thermodynamic states are prescribed. The dispersion relation therefore depends only on four active closure parameters: the feedback sensitivities $k_\lambda$ and $k_\eta$, and the dimensionless delays $\delta_r$ and $\delta_v$. The non-oscillatory evaporation-to-reaction progress ratio, $\kappa = \eta_0/\lambda_0$, is set to unity because the two progress variables are expected to remain of comparable magnitude, and it is not varied in the following analysis.

Figure 10 shows the resulting stability structure for the representative set $k_\lambda = k_\eta = -1$ and $\delta_r = \delta_v = 2$, over $10^{-4} \leq Da \leq 10^4$. The dominant eigenvalue follows a non-monotonic path in the $(\omega, \sigma)$ plane. At small $Da$, evaporation is fast relative to reaction, so the evaporative relaxation remains closely tied to the perturbed shock state. The mode retains a finite frequency, but $\sigma < 0$, so the perturbation will decay back to the non-oscillatory state. For $Da = O(0.1 - 1)$, evaporation and reaction evolve on comparable timescales. The two delayed responses then maintain a finite phase offset, and the growth rate crosses the neutral axis at a finite frequency. Within the interval bounded by the two star-shaped markers in figure 10, $\sigma > 0$ and $\omega \neq 0$, identifying a self-excited oscillatory mode.

At larger $Da$, the unstable window closes and the branch returns to $\sigma < 0$. The frequency decreases progressively and tends toward zero in the slow-evaporation limit, where the chemical response adjusts rapidly to the slowly supplied vapour and the two-delay phase relation again collapses. The regime classification in Figure 10(c), therefore, reduces the calculation to a simple conclusion: both asymptotic limits are stable, while self-excitation occurs only when evaporation and reaction operate on comparable timescales, distinguishing the two-phase detonation from its single-phase counterpart.

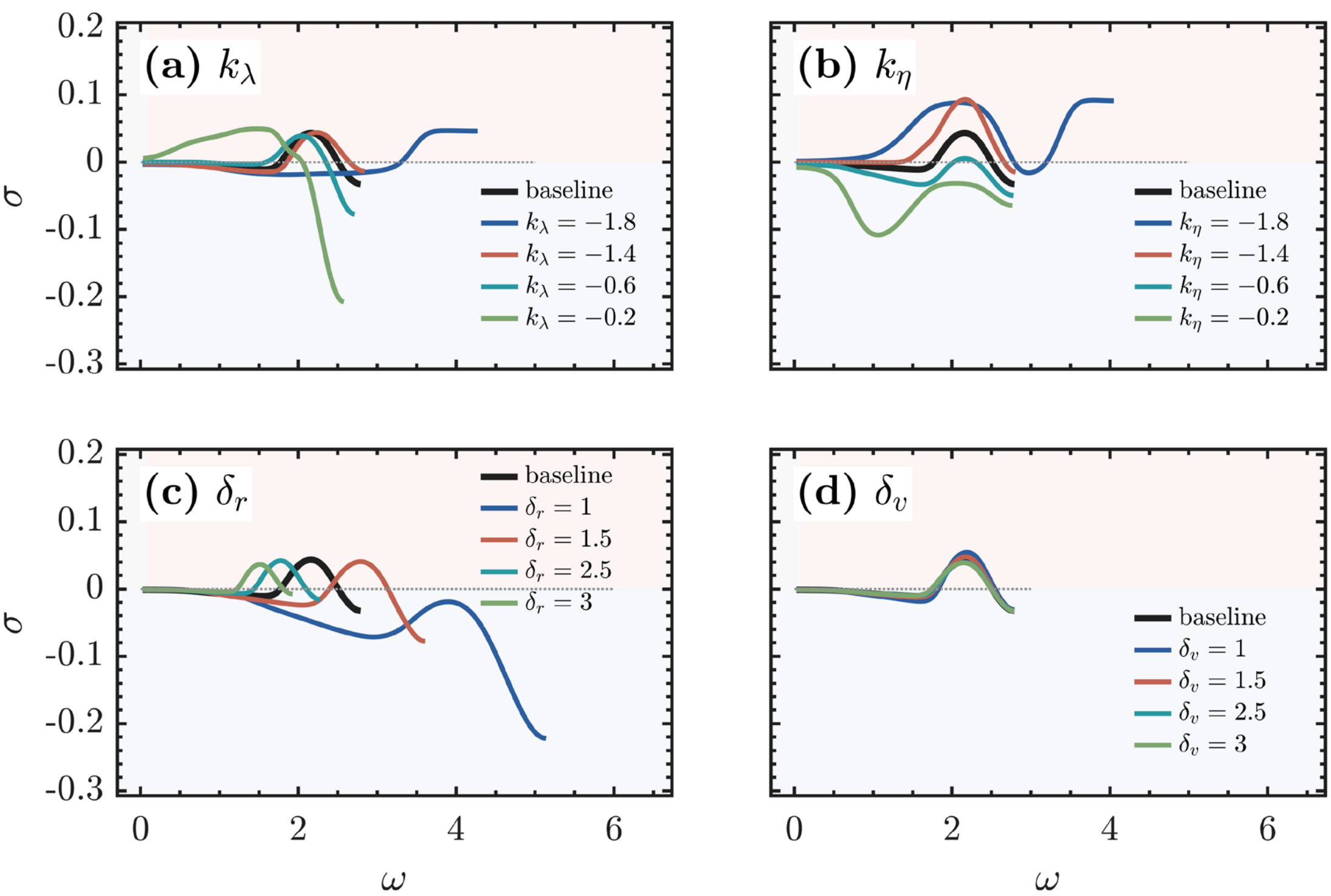

Fig. 11 Deformation of the dominant eigenvalue locus in the $(\omega, \sigma)$ plane under one-at-a-time variation of the four closure parameters. The baseline case is shown in black, and the dotted line denotes the neutral axis $\sigma = 0$. (a) Chemical-side sensitivity, $k_\lambda$. (b) Evaporation-side sensitivity, $k_\eta$. (c) Reaction delay, $\delta_r$ (d) Evaporation delay, $\delta_v$. The corresponding $Da$-resolved curves are given in appendix B.

*3.2.3.2 Influence of parameters on stability transition*

We now vary each closure parameter in turn around the baseline of Section 3.2.3.1 and trace the deformation of the $\sigma$–$\omega$ locus. Figure 11 collects the four parameter sweeps, with the baseline retained in black for reference; the corresponding $\sigma(Da)$ and $\omega(Da)$ curves are deferred to figure B1 of appendix B. The sweeps reveal a clear ordering of the parameters by destabilising strength.

The chemical-side feedback $k_\lambda$ (Figure 11a) acts as a moderate destabiliser. Increasing $|k_\lambda|$ lifts the $\sigma$-peak of the locus toward $\sigma = 0$ and broadens the response in $\omega$, but the locus does not penetrate deeply into the $\sigma > 0$ half-plane: across the explored range of $k_\lambda$, the leading branch remains close to neutrality. Physically, a stronger sensitivity of the chemical equilibrium to a Mach perturbation amplifies the heat-release modulation but does not, on its own, supply the additional phase lead required for sustained self-excitation.

The evaporation-side feedback $k_\eta$ (Figure 11b) is, by contrast, the strongest destabiliser in the family. Increasing $|k_\eta|$ pushes the $\sigma$-peak distinctly above the neutral axis — by a factor of order two relative to the baseline — and concentrates the unstable lobe around the same intermediate $\omega$ as the baseline. This asymmetry between $k_\lambda$ and $k_\eta$ reflects the central physical message of Section 3.2.1: in two-phase ammonia detonations, it is the evaporation channel, not the chemistry, that controls the linear instability, because flash boiling supplies a latent-heat sink with a phase lead unmatched by the gas-phase reaction.

The reaction delay $\delta_r$ (Figure 11c) acts as the principal frequency-tuning parameter. Decreasing $\delta_r$ shifts the $\sigma$-peak toward higher $\omega$ with a modest amplitude change, whereas increasing $\delta_r$ pulls the peak toward lower $\omega$ and broadens the entire locus over a wider $Da$ interval; for $\delta_r = 1$, the locus develops a deep, narrow trough in $\sigma$ near $\omega \sim 1$, anticipating the M-unst. crossings of figure 12. The evaporation delay $\delta_v$ (Figure 11d), in contrast, leaves the $\sigma$–$\omega$ locus nearly invariant: across the sweep range, the four curves collapse onto the baseline, modulated only at the level of the $\sigma$-peak amplitude. $\delta_v$ is therefore the weakest of the four parameters in the linear stability sense — a result consistent with the DDE characteristic equation of Section 3.2.2.5, in which $\delta_v$ enters only multiplicatively through the evaporation transfer function.

Two cross-cutting observations follow. First, the parameters split into a destabilising pair ($k_\eta$, $k_\lambda$— though strongly asymmetric in strength) and a frequency-tuning pair ($\delta_r$, $\delta_v$ — also asymmetric, with $\delta_r$ dominant). Second, almost all four sweeps preserve the qualitative architecture of the baseline locus: a single dominant lobe in ($\sigma$, $\omega$) bracketed by two neutral crossings.

*3.2.3.3 Stability and frequency maps in parameter space*

The $\sigma$–$\omega$ trajectories of Section 3.2.3.2 do not yet specify, at fixed parameter value, the range of $Da$ over which the leading mode is self-excited. To extract the operational stability boundaries we evaluate $\mathcal{D}(s) = 0$ on the two-dimensional grid $(Da, p)$, with $p \in \{k_\lambda, k_\eta, \delta_r, \delta_v\}$ ranged over the same intervals as in Figure 11. Three regimes are distinguished: A stable region (blue) corresponds to $\sigma < 0$ — the front is linearly damped, regardless of the value of $\omega$; A self-excited region (red) is identified by $\sigma > 0$ with $\omega \neq 0$, the linear precursor of a sustained limit-cycle oscillation; A monotonically unstable (grey) regime corresponds to $\sigma > 0$ with $\omega = 0$.

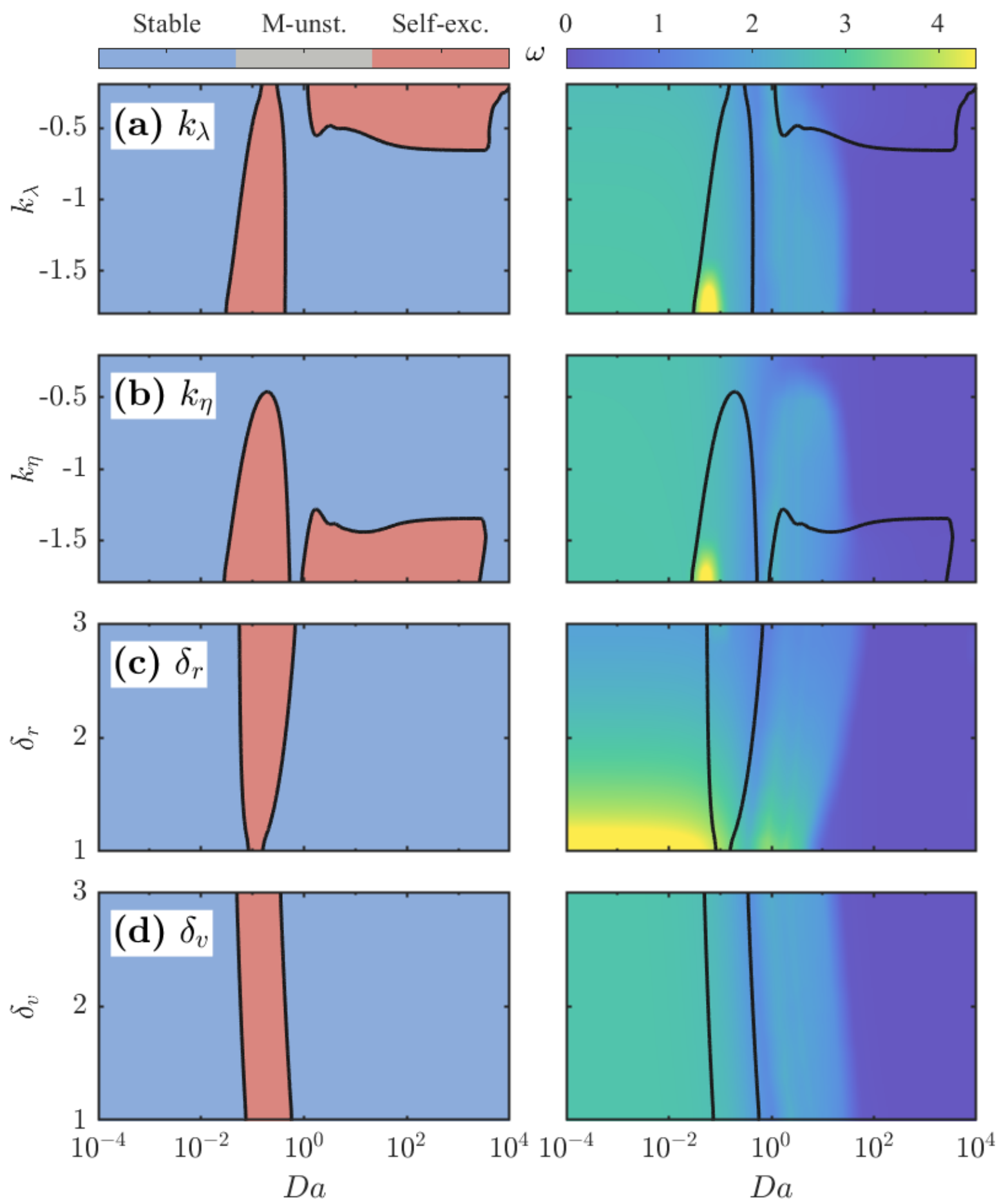

Fig. 12 Stability and frequency maps in the $(Da, p)$ plane with $p \in \{k_\lambda, k_\eta, \delta_r, \delta_v\}$. Each row corresponds to one closure parameter: (a) chemical-side sensitivity, $k_\lambda$; (b) evaporation-side sensitivity, $k_\eta$; (c) reaction delay, $\delta_r$; (d) evaporation delay, $\delta_v$. Left column: stability classification — blue, linearly stable; grey, monotonically unstable; red, self-excited oscillating. Right column: oscillation frequency $\omega$ wherever the dominant branch has a finite imaginary part, with black contours marking the regime boundaries from the left column. Across all four parameters the self-excited region remains anchored to the comparable-timescale interval and the frequency contours track $Da$, confirming that $Da$ is the principal organising parameter while the closure parameters reshape the self-excited tongue and modulate its amplitude.

Figure 12 displays the resulting stability and frequency maps. The destabilising parameters $k_\lambda$ and $k_\eta$ (panels a, b) generate self-excited tongues confined to the comparable-time-scale interval $Da \sim 10^{-1}$–$10^{0}$, in agreement with the asymptotic analysis of Appendix A. The $k_\eta$-tongue (panel b) is the most prominent: a compact ellipse near $(Da, k_\eta) \approx (0.3, -0.5)$ identifies the most strongly self-excited point of the entire family, with a secondary band extending toward larger $Da$ at $k_\eta \lesssim -1.3$. The $k_\lambda$-tongue (panel a) is narrower and remains close to the baseline value of $k_\lambda$, but acquires a long high-$Da$ tail at $|k_\lambda| \lesssim 0.7$ regime, in which the chemistry is barely sensitive to the shock perturbation and the system relaxes into a chemically frozen but evaporatively driven self-excitation. The reaction-delay parameter $\delta_r$(panel c) opens a self-excited tongue with a markedly different geometry: the tongue is widest at $\delta_r \approx 1$ and narrows as $\delta_r$ increases toward 3, consistent with its role as a frequency-tuning rather than amplitude-tuning parameter. The evaporation-delay parameter $\delta_v$ (panel d) generates the narrowest tongue of the family — almost vertical in the $(Da, \delta_v)$ plane.

The right column of figure 12 reports the same $(Da, p)$ planes coloured by the value of $\omega$ wherever the leading branch carries a finite imaginary part, with the regime boundaries from the left column overlaid in black. Two features emerge. First, the maximum of $\omega$ (yellow regions in panels (a)–(d)) coincides with the self-excited tongues of the left column, confirming that the strongest frequency response of the dispersion relation is also the strongest growth-rate response. Second, away from the self-excited tongues, the contours of constant $\omega$ depend almost exclusively on $Da$ and only weakly on the orthogonal parameter: along any horizontal line of fixed $p$, $\omega$ relaxes from $\omega \approx 2.8$ at $Da \ll 1$ to zero at $Da \gg 1$, reproducing the baseline curve of figure 10(b).

The two figures, taken together, establish $Da$ as the principal control parameter of the linear instability. Variations of any of the four closure parameters reshape the $\sigma$–$\omega$ locus and modulate the size and position of the self-excited tongue, but the tongue itself remains anchored to the comparable-time-scale interval $Da \sim 10^{-1}$–$10^{0}$; conversely, a sweep in $Da$ at fixed closure traverses all three regimes within a single decade. The closure parameters thus act as modulators of a $Da$-organised landscape, not as independent control parameters.

*3.2.3.4 Comparison with simulations*

The four closure parameters of the dispersion relation are not fitting constants: each is an order-unity quantity dictated by the thermochemistry of liquid ammonia and by the internal structure of the simulated detonations. The feedback sensitivities $k_\lambda$ and $k_\eta$ measure the differential response of the chemical and evaporation completions to a unit change of the shock Mach number; for a near-Chapman–Jouguet front in an $NH_3/O_2$ mixture the high latent heat ($L_v \approx 1370\ \mathrm{kJ/kg}$) and the steep Arrhenius temperature dependence of $NH_3$ oxidation yield order unity of $k_\lambda$ and $k_\eta$ in absolute values. The dimensionless delays $\delta_r$ and $\delta_v$ are read off the spatial layout of the relaxation zones extracted from Figures 4, 6 and 7: the thin reaction layer immediately behind the shock yields $\delta_r \in [1.5, 3.0]$, while the broader evaporation layer yields $\delta_v \in [0.5, 3.0]$.

We retain every parameter combination $(k_\lambda, k_\eta, \delta_r, \delta_v)$ within these bounds whose dispersion-relation solution simultaneously contains all eight numerical data points within its $\omega$-prediction interval and remains in the self-excited oscillatory half-plane $\sigma > 0$, consistent with the bounded but sustained oscillations observed in the simulations. Figure 13 plots the resulting theoretical response in the $(Da, \omega)$ plane as two nested envelopes: a broader support band spanning the full admissible family, a narrower core band spanning the central majority, the family median (dashed line), and a representative best-fit member (solid blue line). The oscillating frequency for the numerical data here is nondimensionalised as $\omega = 2\pi f \tau_r$. The numerical data are overlaid as circles. The light-grey vertical strip marks the top-three $Da$ window: the three simulation cases lying closest to the theoretical maximum of $\omega$, which in the present figure occurs at $Da \approx 0.5$.

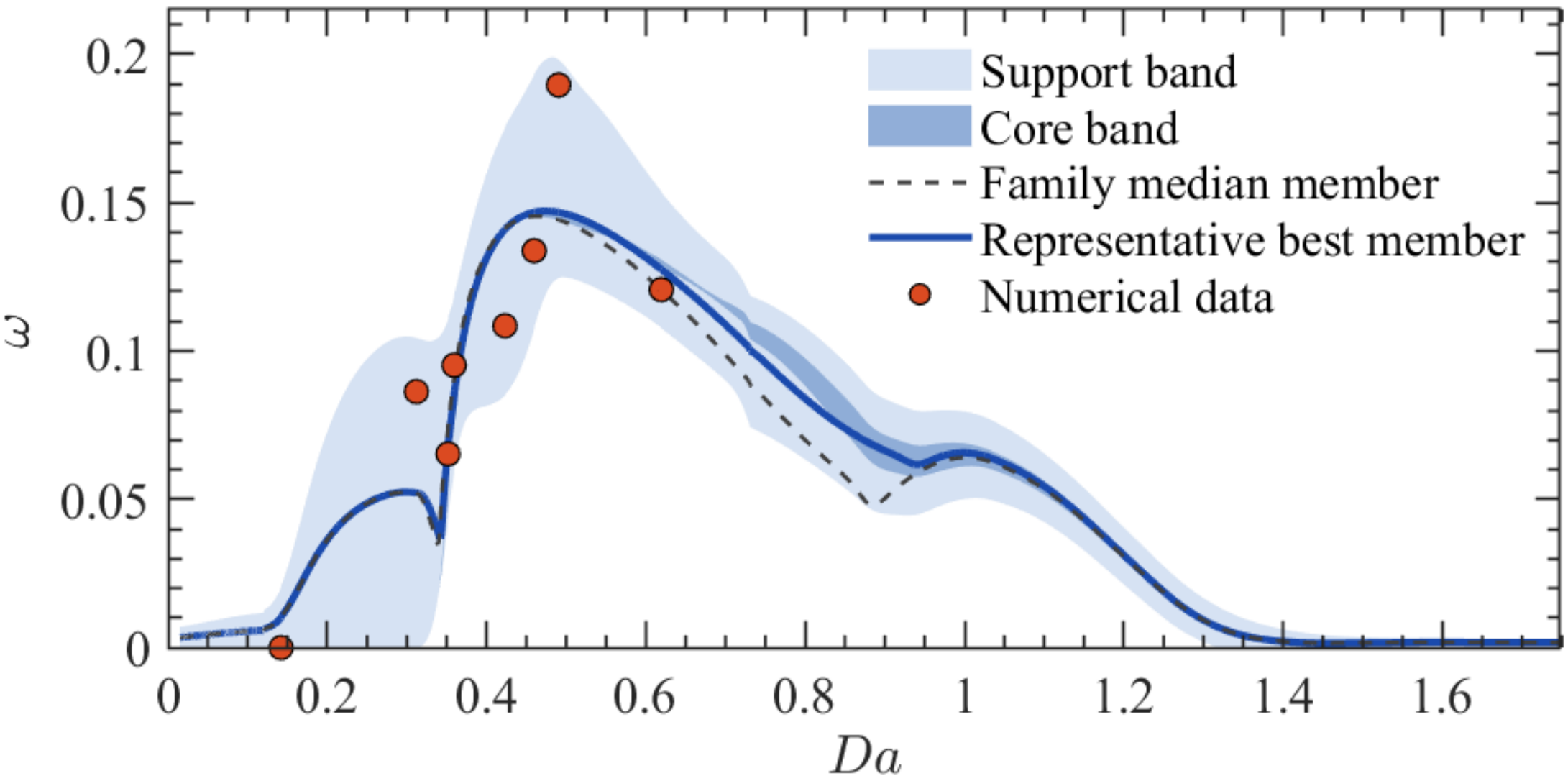

Fig. 13 Comparison between the reduced DDE model and Eulerian–Lagrangian simulations in the ($Da$, $\omega$) plane. The light shaded region denotes the support band obtained from the full admissible parameter family, the darker band denotes the corresponding core band, the dashed curve gives the family median, and the solid blue curve shows one representative member. Symbols denote numerical data extracted from Fig. 8, including both conventional-evaporation and flash-boiling cases. The DDE model captures the non-monotonic frequency selection and envelopes the numerical oscillation frequencies.

All eight numerical points fall within the support band, and seven lie within the narrower core band. The single excursion above the core band, near $Da \simeq 0.49$ and $\omega \simeq 0.19$, remains bracketed by the upper edge of the support envelope and occurs close to the theoretical frequency maximum. The representative best member, corresponding approximately to $k_\lambda \simeq -1.2$, $k_\eta \simeq -0.9$, $\delta_r \simeq 2.1$, and $\delta_v \simeq 1.4$, reproduces both the rapid rise of $\omega$ across the transition range and the broad plateau that follows. This agreement shows that the reduced stability model captures the frequency-selection mechanism observed in the nonlinear simulations. The comparison should not be read as a prediction of oscillation amplitude. Instead, the linear theory identifies the finite-frequency branch, while the full Eulerian–Lagrangian dynamics bound the response through the flash-boiling–chemistry feedback loop identified in Section 3.2.1.

The collapse of the eight simulation results onto a single $Da$-controlled curve, despite the diversity of physical conditions sampled (two evaporation models, two ambient oxygen temperatures and two droplet diameters), validates the central claim of the present analysis: the multiplicity of two-phase ammonia detonation behaviours observed in Section 3.1 is not the manifestation of distinct instability mechanisms but the expression of a single instability governed by the ratio $Da = \tau_v/\tau_r$. In the fast-evaporation limit $\tau_v \ll \tau_r$, droplets vaporise within the induction zone and the front recovers a gas-phase-detonation-like state; in the slow-evaporation limit $\tau_v \gg \tau_r$, droplets traverse the reaction zone essentially unaltered, and the front recovers an inert-shock-like state. Only when $\tau_v \sim \tau_r$ do the two relaxations couple resonantly, and the closed feedback loop between shock strength, latent-heat extraction, and chemical heat release produces the self-sustained oscillations that are the central object of the present study.

## 4. Conclusion

Phase-change kinetics plays a pivotal role of in liquid-ammonia detonation dynamics. By using high-fidelity Eulerian–Lagrangian simulations, we discovered a self-excited oscillatory state behind the leading shock of an ammonia detonation within a droplet-laden region. Parametric studies show that the oscillation is intrinsic to the flash-boiling of ammonia. The dispersed-phase properties modulate rather than initiate the oscillation. Increasing the droplet diameter amplifies the oscillation by enlarging the latent-heat reservoir convected into the post-shock region; the response of oscillating frequency to ambient temperature is non-monotonic, reflecting the competing influences of the post-shock superheat available for flash boiling and the chemical induction time of the gas-phase reaction.

The hypothesized mechanism of the oscillating detonation is a time-delayed competition between latent-heat absorption and chemical heat release. Specifically, a stronger shock (caused by, say, a perturbation) shortens the evaporation length, localises latent-heat adsorption from ambience, cools the gas mixture near the shock, slows down the chemical reaction rates, and pushes heat release downstream, and eventually weakens the front. The weakened shock then spreads evaporation over a longer distance, pushing the radical pool and combustion heat release closer to the front, and then strengthens the front again. The present theory formulates this mechanism by closing a shock-response constraint with delayed evaporation and reaction progress variables. The resulting dispersion relation is expressed by the evaporation Damköhler number, $Da = \tau_v/\tau_r$: both asymptotic limits of $Da \gg 1$ and $Da \ll 1$ are stable, whereas a finite-frequency self-excited window appears for $Da \sim O(1)$. The simulation results collapse onto this $Da$-dependent frequency band, substantiating the theoretical model.

A few extensions of the present work are straightforward. The first is to refine the non-equilibrium boiling closure beyond the Adachi-type correlation, with particular attention to the early-time superheat regime in which the Zuo et al.'s nucleation model differs most from the quasi-steady Abramzon–Sirignano model. The second is to map the sensitivity of the self-excited Hopf oscillating window to additional physico-chemical inputs, in particular the choice of detailed chemical mechanism and the choice of liquid fuel. The last is a normal-mode analysis carried out directly on the 1D two-phase reactive Euler equations, in which the evaporation and reaction source terms appear as integro-differential operators in the post-shock disturbance equations. Such a theoretical approach would retain a transcendental dispersion relation of the same DDE type, but could resolve the spatial structure of the disturbance within the relaxation zone.

**Acknowledgements**

This work was supported by the National Natural Science Foundation of China (Grant No. 52176134 and 12172365). The work at the City University of Hong Kong was additionally supported by grants from the Research Grants Council of the Hong Kong Special Administrative Region, China (Project No. CityU 15222421 and CityU 15218820).

## APPENDIX A

The characteristic equation $\mathcal{D}(s)=0$ in the form (3.17) is examined here in the two distinguished limits $\mathrm{Da}\ll 1$ and $\mathrm{Da}\gg 1$. The neutral root $s=0$ is excluded throughout, and only the principal logarithmic branch is retained, since higher logarithmic sheets correspond to delay harmonics that do not represent the lowest-frequency mode of interest. A dimensional frequency may be assigned to the dominant complex pair of the full finite-delay problem,

$$f_{\mathrm{th}}=\frac{\operatorname{Im} s}{2\pi\,\tau_r},\qquad (A1)$$

but, as shown below, the endpoint asymptotics give real principal roots on the branch-balance locus and therefore do not themselves predict a finite frequency. The finite-frequency mode instead belongs to the comparable-time-scale problem, for which an explicit Hopf slice is recorded first.

### *A.1 An explicit $O(1)$ Hopf oscillating slice*

Set $\mathrm{Da}=1$ and $\delta_r=\delta_v=\delta$ in (3.17), and introduce $A=1-q_r+q_v$ and $B=q_v/(\chi\kappa)-\chi\kappa q_r$. The characteristic equation becomes

$$\mathcal{D}_c(s)=s^2+(1+A)s+A\left(1-\mathrm{e}^{-2s\delta}\right)+Bs\,\mathrm{e}^{-s\delta}=0.\qquad (A2)$$

Looking for a neutral oscillation $s=\mathrm{i}\Omega$, $\Omega>0$, and writing $\theta=\Omega\delta$, the real and imaginary parts of (A2) are

$$-\Omega^2+A(1-\cos 2\theta)+B\Omega\sin\theta=0,\qquad (A3)$$

$$(1+A)\Omega+A\sin 2\theta+B\Omega\cos\theta=0.\qquad (A4)$$

For any $O(1)$ pair $(\Omega,\theta)$ with $\sin\theta\neq 0$, these equations give the exact neutral surface

$$A_*=-1-\Omega\cot\theta,\qquad B_*=\frac{\Omega}{\sin\theta}+2\cos\theta+\frac{2\sin\theta}{\Omega}.\qquad (A5)$$

Hence the finite-delay model contains purely imaginary roots in the comparable-time-scale regime without invoking either endpoint limit.

The branch-balance subcase $A=0$ gives the simpler non-zero spectral equation (3.18). Substitution of $s=\mathrm{i}\Omega$ gives

$$1+B\cos(\Omega\delta)=0,\qquad \Omega-B\sin(\Omega\delta)=0,\qquad (A6)$$

and therefore

$$\Omega=\sqrt{B^2-1},\qquad \delta_n=\frac{\arccos(-1/B)+2\pi n}{\sqrt{B^2-1}},\quad n=0,1,2,\dots,\qquad (A7)$$

provided $B>1$. The principal branch $n=0$ gives the delay in (3.19) and the roots $s_\pm=\pm\mathrm{i}\Omega$. To check that this neutral pair is a genuine Hopf crossing, write $D(s,\delta)=s+1+B\exp(-s\delta)$.

Since

$$\frac{\mathrm{d}s}{\mathrm{d}\delta} = -\frac{D_\delta}{D_s}, \tag{A8}$$

evaluation at $s = \mathrm{i}\Omega$ and $\delta = \delta_0$ gives

$$\left.\frac{\mathrm{dRe}(s)}{\mathrm{d}\delta}\right|_{\delta=\delta_0} = \frac{\Omega^2}{(1+\delta_0)^2 + (\delta_0\Omega)^2} > 0. \tag{A9}$$

Thus increasing the common delay beyond $\delta_0$ sends the conjugate pair into the right half-plane and produces a growing finite-frequency disturbance.

*A.2 Fast-vaporisation limit ($Da \ll 1$)*

In this limit the evaporation relaxation time Da in (3.8) is small. Setting $\varepsilon = \mathrm{Da}$ and writing $s = s_0 + \varepsilon s_1 + O(\varepsilon^2)$, the leading balance of (3.17) gives

$$(1+q_v)\, s_0 + (1 - q_r + q_v)\left(1 - \mathrm{e}^{-s_0\delta_r}\right) - \chi\kappa q_r\, s_0\, \mathrm{e}^{-s_0\delta_r} = 0. \tag{A10}$$

On the branch-balance locus $1 - q_r + q_v \to 0$, and excluding the neutral root $s_0 = 0$, (A10) reduces to $(1+q_v) - \chi\kappa q_r \exp(-s_0\delta_r) = 0$, with principal root

$$s_0 = \frac{1}{\delta_r}\operatorname{Log}\left(\frac{\chi\kappa q_r}{1+q_v}\right). \tag{A11}$$

For positive real gain $\chi\kappa q_r/(1+q_v) > 0$, this principal root is real, and the fast-vaporisation endpoint does not by itself describe a finite-frequency oscillation; the disturbance reduces to a quasi-single-phase shock response with the evaporation delay collapsed.

*A.3 Slow-vaporisation limit ($Da \gg 1$)*

In this limit the evaporation relaxation time Da is large. Setting $z = s\,\mathrm{Da}$ and writing $z = z_0 + \mathrm{Da}^{-1} z_1 + O(\mathrm{Da}^{-2})$, division of (3.17) by Da at leading order gives

$$(1 - q_r)\, z_0 + (1 - q_r + q_v)\left(1 - \mathrm{e}^{-z_0\delta_v}\right) + q_v \chi^{-1}\kappa^{-1}\, z_0\, \mathrm{e}^{-z_0\delta_v} = 0. \tag{A12}$$

On the branch-balance locus $1 - q_r + q_v \to 0$, equivalently $1 - q_r = -q_v$, and excluding the neutral root, (A12) reduces to $\exp(-z_0\delta_v) = \chi\kappa$, with principal root

$$z_0 = -\frac{1}{\delta_v}\operatorname{Log}(\chi\kappa). \tag{A13}$$

For $\chi\kappa > 0$ this principal root is real, and through $s = z/\mathrm{Da}$ the slow-vaporisation endpoint also gives a real principal eigenvalue. The reaction zone has equilibrated to the slowly evolving vapour supply, and the two-phase delay structure of (3.17) no longer survives at leading order.

## APPENDIX B

This appendix collects the $D$a-resolved variations of the growth rate $\sigma$ and oscillation frequency $\omega$ in Figure B1 for the four closure-parameter families discussed in Section 3.2.3.2.

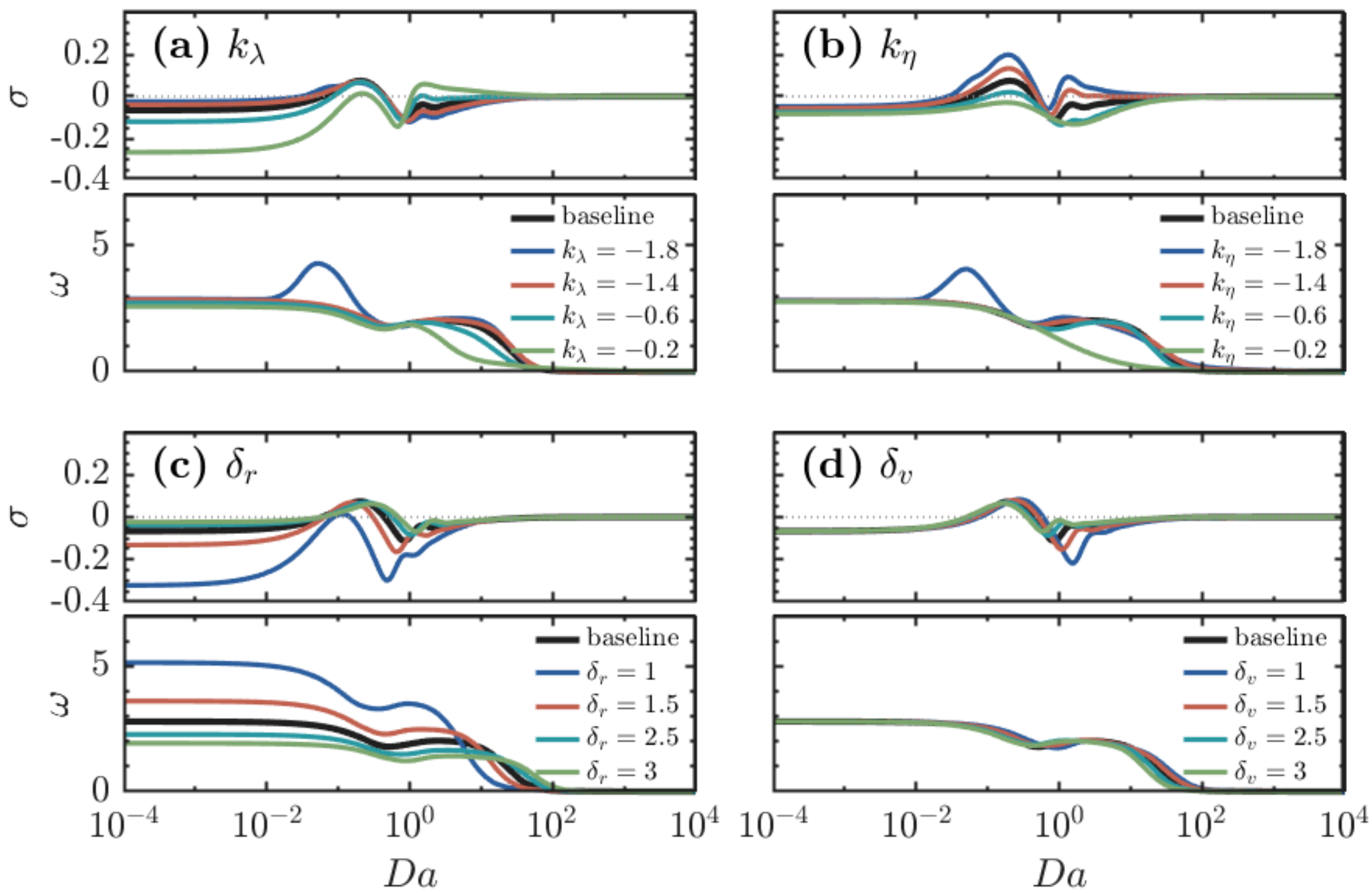

**Fig. B1.** Growth rate and oscillation frequency predicted by the DDE model as functions of $Da$ for the four closure-parameter families introduced in Section 3.2.3.2. The baseline case is shown in black, and the dotted line denotes the neutral axis $\sigma = 0$. (a) Chemical-side sensitivity $k_\lambda$. (b) Evaporation-side sensitivity $k_\eta$. (c) Reaction delay $\delta_r$. (d) Evaporation delay $\delta_v$.

## References:

[1] K. Kailasanath, Review of propulsion applications of detonation waves, AIAA Journal 38 (2000) 1698-1708.

[2] B. Zhang, H.D. Ng, J.H.S. Lee, The critical tube diameter and critical energy for direct initiation of detonation in C2H2/N2O/Ar mixtures, Combustion and Flame 159 (2012) 2944-2953.

[3] Z. Zhang, Y. Liu, C. Wen, Mechanisms of the destabilized Mach reflection of inviscid oblique detonation waves before an expansion corner, Journal of Fluid Mechanics 940 (2022) A29.

[4] G.D. Roy, S.M. Frolov, A.A. Borisov, D.W. Netzer, Pulse detonation propulsion: challenges, current status, and future perspective, Progress in Energy and Combustion Science 30 (2004) 545-672.

[5] M. Hishida, T. Fujiwara, P. Wolanski, Fundamentals of rotating detonations, Shock Waves 19 (2009) 1-10.

[6] Z. Jiang, Standing oblique detonation for hypersonic propulsion: A review, Progress in Aerospace Sciences 143 (2023) 100955.

[7] R. Zhu, X. Fang, C. Xu, M. Zhao, H. Zhang, M. Davy, Pulsating one-dimensional detonation in ammonia-hydrogen–air mixtures, International Journal of Hydrogen Energy 47 (2022) 21517-21536.

[8] Z. An, J. Xing, R. Kurose, Recent progresses in research on liquid ammonia spray and combustion: A review, Applications in Energy and Combustion Science 20 (2024) 100293.

[9] M. Short, I.I. Anguelova, T.D. Aslam, J.B. Bdzil, A.K. Henrick, G.J. Sharpe, Stability of detonations for an idealized condensed-phase model, Journal of Fluid Mechanics 595 (2008) 45-82.

[10] H. Watanabe, A. Matsuo, A. Chinnayya, K. Matsuoka, A. Kawasaki, J. Kasahara, Numerical analysis of the mean structure of gaseous detonation with dilute water spray, Journal of Fluid Mechanics 887 (2020) A4.

[11] K. Kailasanath, Liquid-fueled detonations in tubes, Journal of Propulsion and Power 22 (2006) 1261-1268.

[12] W.C. Davis, W. Fickett, Detonation – Theory and Experiment, (1969).

[13] J.H.S. Lee, The Detonation Phenomenon, Cambridge University Press, Cambridge, 2008.

[14] J.J. Erpenbeck, Stability of Idealized One-Reaction Detonations, The Physics of Fluids 7 (1964) 684-696.

[15] J.J. Erpenbeck, Stability of Steady-State Equilibrium Detonations, The Physics of Fluids 5 (1962) 604-614.

[16] A. Bourlioux, A.J. Majda, V. Roytburd, Theoretical and Numerical Structure for Unstable One-Dimensional Detonations, SIAM Journal on Applied Mathematics 51 (1991) 303-343.

[17] H.I. Lee, D.S. Stewart, Calculation of linear detonation instability: one-dimensional instability of plane detonation, Journal of Fluid Mechanics 216 (1990) 103-132.

[18] G.J. Sharpe, Linear stability of idealized detonations, Proceedings of the Royal Society A: Mathematical, Physical and Engineering Sciences 453 (1997) 2603-2625.

[19] G.E. Abouseif, T.Y. Toong, Theory of unstable one-dimensional detonations, Combustion and Flame 45 (1982) 67-94.

[20] P. Clavin, L. He, Stability and nonlinear dynamics of one-dimensional overdriven detonations in gases, Journal of Fluid Mechanics 306 (1996) 353-378.

[21] P. Clavin, L. He, F.A. Williams, Multidimensional stability analysis of overdriven gaseous detonations, Physics of Fluids 9 (1997) 3764-3785.

[22] R. Daou, P. Clavin, Instability threshold of gaseous detonations, Journal of Fluid Mechanics 482 (2003) 181-206.

[23] A.R. Kasimov, L.M. Faria, R.R. Rosales, Model for shock wave chaos, Physical Review Letters 110 (2013) 104104.
[24] M. Short, J.J. Quirk, On the nonlinear stability and detonability limit of a detonation wave for a model three-step chain-branching reaction, Journal of Fluid Mechanics 339 (1997) 89-119.
[25] A.R. Kasimov, D.S. Stewart, Spinning instability of gaseous detonations, Journal of Fluid Mechanics 466 (2002) 179-203.
[26] H. Ng, A. Higgins, C. Kiyanda, M. Radulescu, J. Lee, K. Bates, N. Nikiforakis, Nonlinear dynamics and chaos analysis of one-dimensional pulsating detonations, Combustion Theory and Modelling 9 (2005) 159-170.
[27] H.A. Abderrahmane, F. Paquet, H.D. Ng, Applying nonlinear dynamic theory to one-dimensional pulsating detonations, Combustion Theory and Modelling 15 (2011) 205-225.
[28] C.á. Romick, T.D. Aslam, J. Powers, The effect of diffusion on the dynamics of unsteady detonations, Journal of Fluid Mechanics 699 (2012) 453-464.
[29] M.I. Radulescu, G.J. Sharpe, C.K. Law, J.H.S. Lee, The hydrodynamic structure of unstable cellular detonations, Journal of Fluid Mechanics 580 (2007) 31-81.
[30] A. Sow, A. Chinnayya, A. Hadjadj, Mean structure of one-dimensional unstable detonations with friction, Journal of Fluid Mechanics 743 (2014) 503-533.
[31] F. Zhang, J.H.S. Lee, Friction-induced oscillatory behaviour of one-dimensional detonations, Proceedings of the Royal Society of London. Series A: Mathematical and Physical Sciences 446 (1994) 87-105.
[32] F. Zhang, R. Chue, J. Lee, R. Klein, A nonlinear oscillator concept for one-dimensional pulsating detonations, Shock Waves 8 (1998) 351-359.
[33] J.-P. Dionne, H.D. Ng, J.H. Lee, Transient development of friction-induced low-velocity detonations, Proceedings of the Combustion Institute 28 (2000) 645-651.
[34] W. Han, C. Wang, C.K. Law, Pulsation in one-dimensional H2–O2 detonation with detailed reaction mechanism, Combustion and Flame 200 (2019) 242-261.
[35] K. Mazaheri, Y. Mahmoudi, M.I. Radulescu, Diffusion and hydrodynamic instabilities in gaseous detonations, Combustion and Flame 159 (2012) 2138-2154.
[36] Z. Weng, R. Mével, Linear and non-linear stability of gaseous detonation at elevated pressure, Combustion and Flame 262 (2024) 113361.
[37] T. Lu, C.K. Law, Heterogeneous Effects in the Propagation and Quenching of Spray Detonations, Journal of Propulsion and Power 20 (2004) 820-827.
[38] N. Tricard, X. Zhao, One dimensional modeling of spray detonations considering loss effects, AIAA SCITECH 2022 Forum2022.
[39] D. Martínez-Ruiz, On the structure of steady one-dimensional liquid-fueled detonations, Physics of Fluids 35 (2023).
[40] R. Hernández-Sánchez, C. Huete, D. Martínez-Ruiz, Pathological detonations in mono-disperse spray media, Proceedings of the Combustion Institute 40 (2024) 105505.
[41] S. Cheatham, K. Kailasanath, Numerical modelling of liquid-fuelled detonations in tubes, Combustion Theory and Modelling 9 (2005) 23-48.
[42] H. Watanabe, A. Matsuo, K. Matsuoka, A. Kawasaki, J. Kasahara, Numerical investigation on propagation behavior of gaseous detonation in water spray, Proceedings of the Combustion Institute 37 (2019) 3617-3626.
[43] Q. Meng, M. Zhao, Y. Xu, L. Zhang, H. Zhang, Structure and dynamics of spray detonation in n-

heptane droplet/vapor/air mixtures, Combustion and Flame 249 (2023) 112603.
[44] D.A. Schwer, Multi-dimensional Simulations of Liquid-Fueled JP10/Oxygen Detonations, AIAA Propulsion and Energy 2019 Forum.
[45] Z. Huang, Y. Xu, S. Li, Q. Meng, H. Zhang, Detailed simulations of one-dimensional detonation propagation in dilute n-heptane spray and air mixtures, Physics of Fluids 35 (2023).
[46] R. Zhu, M. Zhao, H. Zhang, Numerical simulation of flame acceleration and deflagration-to-detonation transition in ammonia-hydrogen–oxygen mixtures, International Journal of Hydrogen Energy 46 (2021) 1273-1287.
[47] R. Zhu, G. Li, F. Leach, M. Davy, Numerical simulation of detonation propagation and extinction in two-phase gas-droplet ammonia fuel, International Journal of Hydrogen Energy 90 (2024) 218-229.
[48] J. Sun, Y. Wang, S.M. Shaik, H. Zhang, Numerical investigation of detonation initiation and propagation in non-uniformly cracked ammonia and air mixtures, Combustion and Flame 281 (2025) 114439.
[49] J. Sun, S.M. Shaik, V. Bo Nguyen, H. Zhang, Detonation chemistry and propagation characteristics in partially cracked ammonia, Proceedings of the Combustion Institute 41 (2025) 105910.
[50] F. Veiga-López, R. Mével, Detonation properties and nitrogen oxide production in ammonia–hydrogen–air mixtures, Fuel 370 (2024) 131794.
[51] Y. Song, H. Hashemi, J.M. Christensen, C. Zou, P. Marshall, P. Glarborg, Ammonia oxidation at high pressure and intermediate temperatures, Fuel 181 (2016) 358-365.
[52] B. Abramzon, W.A. Sirignano, Droplet vaporization model for spray combustion calculations, International Journal of Heat and Mass Transfer 32 (1989) 1605-1618.
[53] R.H. Perry, D.W. Green, J.O. Maloney, Perry's Chemical Engineers' Handbook, 2007.
[54] B. Zuo, A.M. Gomes, C.J. Rutland, Modelling superheated fuel sprays and vaproization, International Journal of Engine Research 1 (2000) 321-336.
[55] M. Adachi, V.G. Mcdonell, D. Tanaka, J. Senda, H. Fujimoto, Characterization of fuel vapor concentration inside a flash boiling spray, (1998).
[56] Z. Huang, H. Wang, K. Luo, J. Fan, Large eddy simulation investigation of ammonia spray characteristics under flash and non-flash boiling conditions, Applications in Energy and Combustion Science 16 (2023) 100220.
[57] Z. An, J. Xing, R. Kurose, Numerical study on the phase change and spray characteristics of liquid ammonia flash spray, Fuel 345 (2023) 128229.
[58] A.B. Liu, D. Mather, R.D. Reitz, Modeling the effects of drop drag and breakup on fuel sprays, SAE Transactions, (1993) 83-95.
[59] R. Zhu, Z. Huang, C. Xu, B. Wu, M. Davy, Integration and validation of some modules for modelling of high-speed chemically reactive flows in two-phase gas-droplet mixtures, Computers & Fluids 277 (2024) 106282.
[60] W.E. Ranz, Evaporation from drops: Part I, Chemical Engineering Progress 48 (1952) 141.
[61] Z. Huang, M. Zhao, Y. Xu, G. Li, H. Zhang, Eulerian-Lagrangian modelling of detonative combustion in two-phase gas-droplet mixtures with OpenFOAM: Validations and verifications, Fuel 286 (2021) 119402.
[62] Z. Ren, B. Wang, G. Xiang, L. Zheng, Effect of the multiphase composition in a premixed fuel–air stream on wedge-induced oblique detonation stabilisation, Journal of Fluid Mechanics 846 (2018) 411-427.
[63] H. Chen, M. Zhao, H. Qiu, Y. Zhu, Implementation and verification of an OpenFOAM solver for

gas-droplet two-phase detonation combustion, Physics of Fluids 36 (2024).
[64] M. Pilch, C. Erdman, Use of breakup time data and velocity history data to predict the maximum size of stable fragments for acceleration-induced breakup of a liquid drop, International Journal of Multiphase Flow 13 (1987) 741-757.
[65] H. Teng, C. Tian, P. Yang, M. Zhao, Effect of droplet diameter on oblique detonations with partially pre-vaporized n–heptane sprays, Combustion and Flame 258 (2023) 113062.
[66] C. Tian, H. Teng, B. Shi, P. Yang, K. Wang, M. Zhao, Propagation instabilities of the oblique detonation wave in partially prevaporized n-heptane sprays, Journal of Fluid Mechanics 984 (2024) A16.
[67] H. Guo, Y. Sun, R. Zhu, S. Wang, M. Zhao, B. Shi, X. Hou, Inhibition of the oblique detonation wave detachment in two-phase n-heptane/air mixtures, Combustion and Flame 272 (2025) 113843.
[68] Q. Meng, Z. Wang, On the direct initiation in n-heptane mists considering droplet fragmentation, Physics of Fluids 37 (2025).
[69] C.T. Crowe, M.P. Sharma, D.E. Stock, The Particle-Source-In Cell (PSI-CELL) Model for Gas-Droplet Flows, Journal of Fluids Engineering 99 (1977) 325-332.
[70] W. Wang, M. Yang, Z. Hu, P. Zhang, A dynamic droplet breakup model for Eulerian-Lagrangian simulation of liquid-fueled detonation, Aerospace Science and Technology 151 (2024) 109271.
[71] W. Wang, Z. Hu, P. Zhang, Liquid-fueled oblique detonation waves induced by reactive and non-reactive transverse liquid jets, Aerospace Science and Technology 168 (2026) 111293.
[72] W. Wang, Z. Hu, P. Zhang, Computational investigation on the formation of liquid-fueled oblique detonation waves, Combustion and Flame 271 (2025) 113839.
[73] H.S. Tsien, Servo-stabilization of combustion in rocket motors, Journal of the American Rocket Society 22 (1952) 256-263.
[74] L. Crocco, S.-I. Cheng, Theory of Combustion Instability in Liquid Propellant Rocket Motors, AGARDograph No. 8, Butterworths Scientific Publications, London, 1956.
[75] G.B. Whitham, On the propagation of shock waves through regions of non-uniform area or flow, Journal of Fluid Mechanics 4 (1958) 337-360.
[76] C. Wang, G. Xiang, Z. Jiang, Theoretical approach to one-dimensional detonation instability, Applied Mathematics and Mechanics 37 (2016) 1231-1238.